**Generation and non-thermal annealing of dislocations and atomic-scale roughness by nonlinear surface acoustic waves**

Yuan Xu, Maxim V. Shugaev, and Leonid V. Zhigilei*

Department of Materials Science and Engineering, University of Virginia,
395 McCormick Rd, Charlottesville, VA, 22904, USA

## Abstract

Surface acoustic waves (SAWs) are widely used to probe near-surface material properties and microstructure, but their potential to drive permanent structural modification remains largely unexplored. Large-scale molecular dynamics simulations are used to reveal the mechanisms of defect generation and evolution induced by nonlinear Rayleigh waves in Pt. Strong crystallographic dependence of the nonlinear evolution is observed. SAWs propagating on the Pt(111) surface undergo rapid nonlinear sharpening and shock formation, accompanied by localization of elastic energy and increasing stresses near the surface, whereas much weaker sharpening on Pt(001) allows dissipation to compete effectively with shock formation. For sufficiently strong SAWs on Pt(111), Shockley partial dislocations are emitted from the surface when the resolved shear stress approaches the theoretical shear strength. The highly localized and rapidly changing stress field at the shock front activates multiple slip systems and promotes cross-slip and reactions between Shockley partials, producing complex dislocation structures containing sessile stair-rod segments. Subsequent interactions with the propagating SAW activate dislocation sources on slip systems for which the stress is insufficient for homogeneous nucleation, leading, in one case, to the sequential emission of Shockley partials and formation of a subsurface twinned domain. Repeated SAW passages also progressively reduce the dislocation density, demonstrating SAW-driven nonthermal annealing. Interactions of dislocations with the free surface generate atomic-scale steps, terraces, and other under-coordinated configurations, providing a microscopic pathway for permanent surface modification that may contribute to the acoustically enhanced catalytic activity observed on Pt surfaces. These results establish nonlinear SAWs as drivers of nonthermal defect generation, restructuring, and annealing, opening opportunities for acoustic processing of near-surface microstructure.

* Corresponding author, E-mail address: lz2n@virginia.edu (L.V. Zhigilei).

## Graphical Abstract

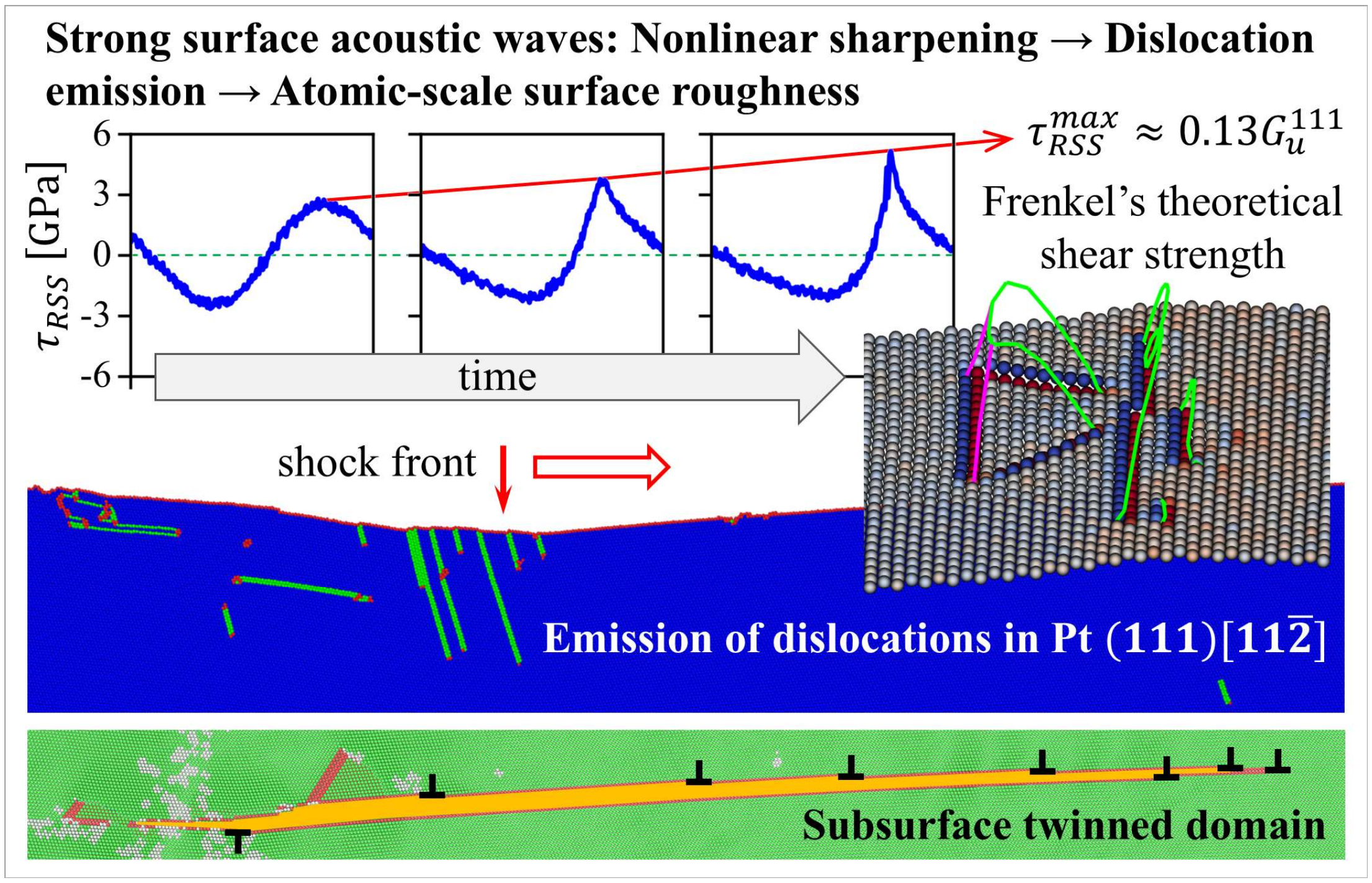

## 1. Introduction

Surface acoustic waves (SAWs), or Rayleigh waves, are elastic waves that propagate along a solid surface while their amplitudes decay exponentially with depth into the material [1-3]. This confinement of acoustic energy to a shallow surface layer, combined with the ability of SAWs to propagate over long distances with relatively low attenuation, gives SAWs a unique combination of sensitivity to near-surface structure and efficient energy transport. As a result, SAWs are widely used for nondestructive characterization of elastic properties and near-surface microstructure [4-6]. In particular, optical generation and detection of SAWs constitute the basis of several materials characterization techniques, including laser ultrasonics [7,8] for detection of subsurface microcracks and other heterogeneities, spatially resolved acoustic spectroscopy [9] for mapping crystallographic orientation and elastic constants in polycrystalline materials, and transient grating spectroscopy [10-13], where coherent acoustic waves are used to probe microstructure evolution, defect accumulation, and transport phenomena on sub-microsecond timescales and with tunable depth sensitivity.

Beyond probing material properties, SAWs can also serve as efficient actuators capable of coupling acoustic energy to surface and interfacial processes. This capability underlies applications involving microscale manipulation of fluids, droplets, and particles in microfluidic systems [14-16], atomization of liquid samples for mass spectrometry analysis [17-19], surface cleaning through removal of submicron particles [20,21], enhancement of surface diffusion of adsorbates and nanoscale clusters [22-25], and modulation of optical response of plasmonic nanostructures [26]. The ability of SAWs to modify the material structure itself, however, has received comparatively little attention. This is despite demonstrations that strong SAWs can facilitate crystallization in amorphous materials [27], induce localized material damage [28,29], or produce arrays of cracks in brittle materials [30-32]. Moreover, surface roughening associated with the SAW-induced generation of subsurface crystal defects has been proposed [33] as a possible mechanism responsible for the persistent observations of substantial acoustic enhancement of the rates and selectivity of heterogeneous catalytic reactions on metal surfaces [34-40]. These observations clearly demonstrate that acoustic waves can drive irreversible structural evolution. Yet they originate from a relatively small number of studies and have not led to a mechanistic understanding of how acoustic energy can be harnessed for controlled microstructural modification and materials processing.

A growing body of evidence indicates that nonlinear evolution of strong SAWs during propagation plays an important role in many of these phenomena. Unlike bulk acoustic waves, nonlinear SAWs exhibit a distinctly nonlocal character in which continuous generation of higher-frequency harmonics progressively sharpens the wave profile, concentrates acoustic energy in the near-surface region, and ultimately produces steep fronts and characteristic cusps in the strain and particle velocity profiles [33-43]. As a result, the transient strains and stresses in the near-surface region can substantially exceed those of the initially generated wave and may approach the theoretical strengths required for defect nucleation and fracture. In brittle materials, this phenomenon has enabled measurements of the intrinsic dynamic fracture strength through the generation of microcracks by nonlinear SAWs [30-32]. In ductile materials, the same mechanism may drive dislocation emission from surfaces at stresses approaching their theoretical limits and may provide a route for controlled, nonthermal generation and evolution of defect structures in a shallow surface region. The resulting dislocation structures and associated atomic-scale surface roughening may in turn open up new opportunities for tailoring the mechanical, chemical, and catalytic properties of material surfaces.

Despite the scientific significance and potential technological relevance of nonlinear SAW-driven structural evolution, the relationships between nonlinear sharpening of SAWs, dissipation of acoustic energy, generation, evolution, and annealing of subsurface defect structures, and the development of atomic-scale surface roughness remain poorly understood. In particular, it is not known how the increasing localization of acoustic energy near the surface affects the onset of dislocation nucleation, how the generated dislocation structures evolve under repeated interactions with the propagating wave, or how the evolution of subsurface defects gives rise to atomic-scale surface roughness. Answering these questions requires a theoretical framework capable of simultaneously describing nonlinear wave propagation, energy dissipation, and the atomistic mechanisms responsible for defect nucleation, evolution, and annihilation.

Molecular dynamics (MD) simulations are particularly well suited for addressing these challenges [22,23,33,44,45]. Unlike continuum descriptions that rely on approximate constitutive relations of finite order and phenomenological representations of defect dynamics, MD simulations naturally incorporate the full anharmonicity of interatomic interactions underlying nonlinear frequency up-conversion and wave sharpening while enabling direct observation of defect nucleation, migration, annihilation, and interactions with propagating acoustic waves.

Moreover, atomistic simulations provide direct access to the mechanisms of acoustic energy dissipation and permit simultaneous analysis of nonlinear wave evolution, defect dynamics, and atomic-scale surface modifications.

In this work, we employ large-scale MD simulations to investigate the coupling between nonlinear SAW sharpening and the generation and evolution of defect structures in the near-surface region of a crystalline metal. The analysis focuses on the interplay between acoustic energy localization and dissipation, subsurface defect dynamics, and atomic-scale changes in surface structure. The simulations reveal how progressive localization of acoustic energy near the surface can drive dislocation emission at transient stress levels approaching the theoretical strength of the material and how repeated interactions between the propagating wave and the generated defects govern the subsequent evolution and nonthermal annealing of subsurface dislocation structures, as well as the development of atomic-scale surface roughness. By elucidating the elementary mechanisms responsible for SAW-induced structural evolution, this work demonstrates how nonlinear SAWs can act not only as probes of near-surface microstructure but also as drivers of its nonthermal modification, providing a mechanistic foundation for nonlinear SAW-based approaches to surface processing and new insights into material behavior under extreme dynamic loading conditions.

## 2. MD simulation of SAWs on Pt substrates

The generation of subsurface crystal defects and the associated atomic-scale surface roughness is investigated using MD simulations of SAW propagation on Pt substrates. The choice of Pt is motivated by experimental observations of acoustic enhancement of catalytic activity of Pt surfaces [35-37,40] and the hypothesis that this enhancement may be related to SAW-induced generation of crystal defects [33,40].

The interatomic interactions in Pt are described by an embedded atom method (EAM) potential fitted in Ref. [46] to the results of density functional theory (DFT) calculations. The 0 K density and three independent elastic constants reported for this EAM potential are $\rho = 20.62\ \mathrm{g/cm^3}$, $C_{11} = 264.348$ GPa, $C_{12} = 221.931$ GPa, and $C_{44} = 68.782$ GPa [46]. Since the simulations of SAW propagation are performed at room temperature, the finite-temperature deformation method [47,48] is used to calculate the equilibrium density and elastic constants predicted by the potential at 300 K. The resulting values are $\rho = 20.451\ \mathrm{g/cm^3}$, $C_{11} = 262.7$ GPa, $C_{12} = 217.7$ GPa, and $C_{44} = 70.0$ GPa.

The MD simulations of SAW propagation on Pt substrates are performed using the "computational synchrotron" approach developed in Ref. [33] to explore nonlinear effects during free propagation of SAWs. This approach takes advantage of periodic boundary conditions applied along the direction of wave propagation, allowing the SAW to repeatedly traverse the computational system while its profile evolves due to the nonlinear elastic response of the substrate. The repeated circulation of the wave can be continued after the onset of irreversible structural changes induced by the shock wave formed through nonlinear sharpening of the SAW profile [30-33], thus enabling investigation of the interactions between the propagating SAW and subsurface defects generated during its previous passages. Alternatively, following the onset of SAW-induced irreversible structural changes, the periodic boundary condition along the propagation direction can be removed and the computational system extended by adding pristine material ahead of the propagating wave.

To explore the dependence of nonlinear Rayleigh wave evolution on surface orientation and propagation direction, two representative configurations are considered for face-centered cubic (fcc) Pt: SAWs propagating along the $[100]$ direction on the $(001)$ surface and along the $[11\bar{2}]$ direction on the $(111)$ surface. Prior to SAW generation, the Pt substrates are equilibrated at 300 K, with the lattice parameters chosen to ensure zero normal stresses in the systems. The initial SAWs are generated by adding the atomic displacements and velocities obtained from analytical solutions of the wave equation for harmonic SAWs in a cubic crystal to the instantaneous positions and velocities of atoms in the thermally equilibrated substrates. For a wave propagating along the [100] direction on the (001) surface, the solution based on the analysis of Rayleigh wave propagation in cubic crystals [49] is provided in Ref. [33]. For a wave propagating along the $[11\bar{2}]$ direction on the $(111)$ surface, the corresponding solution has not been reported in the literature and is therefore provided in Appendix A.

Note that the analytical solutions for Rayleigh waves are obtained within linear elasticity using the elastic constants predicted by the EAM potential and listed above and therefore do not include nonlinear corrections that can become significant at the large wave amplitudes considered in the present work. As a result, the initial displacement and velocity fields imposed at the start of the simulations deviate from the actual wave profiles corresponding to the nonlinear elastic response of the material, leading to the excitation of weak but noticeable spurious bulk waves. These waves propagate away from the surface into the bulk of the thick computational systems, gradually dissipate, and have no discernible effect on the results reported in this work.

The dimensions of the computational systems and the boundary conditions are defined by the specific characteristics of SAWs and the requirements of the "computational synchrotron" approach. The SAWs investigated in the present study are continuous, periodic, unidirectional Rayleigh waves propagating along the *x*-axis, with the *z*-axis oriented normal to the surface. The "computational synchrotron" approach discussed above requires periodic boundary conditions along the direction of wave propagation, and the minimum system length in this direction, $L_x$, should be equal to the wavelength λ of the fundamental harmonic of the generated wave [33]. Periodic boundary conditions are also applied in the *y* direction, parallel to the wavefront of the propagating SAW. The system size in this direction, $L_y$, is chosen as a compromise between the need for a realistic representation of SAW-induced generation and evolution of crystal defects and the computational cost of the simulations.

In the *z* direction, normal to the surface, free and rigid boundary conditions are applied at the top and bottom surfaces of the substrate, respectively. As demonstrated in Ref. [33], the rigid boundary condition prevents dynamic coupling between the top and bottom surfaces of the substrate but can introduce artificial frequency dispersion during wave propagation if the depth of the system represented in the MD simulation is insufficient. Moreover, the system depth required for a realistic representation of SAW propagation on a bulk substrate strongly depends on the crystallographic orientation of the surface and the wave propagation direction. The importance of a sufficient system depth, $L_z$, is illustrated by the depth dependence of one of the stress components, $\sigma_{xx}$, calculated for SAWs with λ = 244 nm propagating along the $[11\bar{2}]$ direction on the $(111)$ surface (Fig.1a) and with λ = 240 nm propagating along the $[100]$ direction on the $(001)$ surface (Fig.1b). In both cases, the stress field extends to depths on the micrometer scale, with a substantially deeper stress distribution observed for the $(111)$ surface.

Based on the considerations discussed above, the MD computational systems have dimensions of 240 nm × 16 nm × 1600 nm (about 384 million atoms) for SAWs propagating along the $[100]$ direction on the $(001)$ surface and 244 nm × 9 nm × 3100 nm (about 432 million atoms) for SAWs propagating along the $[11\bar{2}]$ direction on the $(111)$ surface. The strength of the waves is characterized by the initial strain amplitude, $\varepsilon_{xx}^{0}$, defined as the maximum surface value of $\varepsilon_{xx}$ at the time the SAW is imposed. The strain is calculated from Eq. (A10) in Appendix A as $\varepsilon_{xx} = \partial u_1/\partial x_1$.

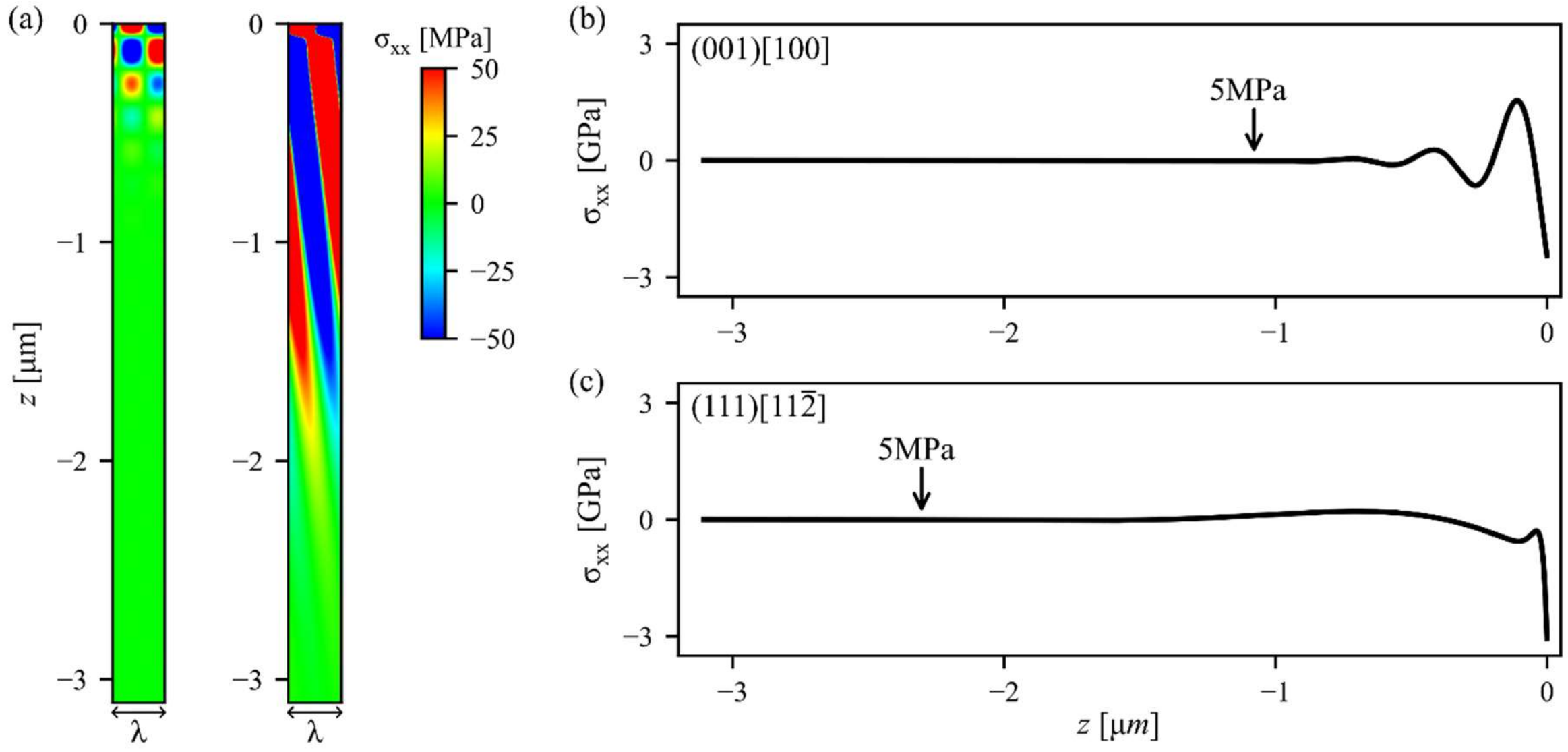

**Fig. 1.** Distributions of the normal stress $\sigma_{xx}$ generated by initial harmonic SAWs in Pt substrates. The two-dimensional distributions of $\sigma_{xx}$ in the *x*–*z* plane are shown in (a) for the $(001)[100]$ SAW (left panel) and $(111)[11\bar{2}]$ SAW (right panel). The corresponding depth profiles for *x* = 0 are shown for a SAW with λ = 240 nm propagating along the $[100]$ direction on the $(001)$ surface in (b) and a SAW with λ = 244 nm propagating along the $[11\bar{2}]$ direction on the $(111)$ surface in (c). The initial strain amplitude is $\varepsilon_{xx}^{0} = 4\%$ in both cases. The arrows in (b) and (c) mark the depths below which $|\sigma_{xx}|$ does not exceed 5 MPa.

For the $(001)$ substrate, SAWs with initial strain amplitudes of 0.5%, 1%, 2%, 4%, 6%, and 8% are simulated. Most simulations are performed for 2 ns, except for the $\varepsilon_{xx}^{0} = 4\%$ and $\varepsilon_{xx}^{0} = 6\%$ cases, which are extended to 4 and 3.2 ns, respectively. For the $(111)$ substrate, SAWs with initial strain amplitudes of 0.5%, 1%, 1.5%, 2%, 2.5%, 3%, and 4% are simulated. Most simulations are performed for 2 ns, except for the $\varepsilon_{xx}^{0} = 1\%$ case, which is extended to 4 ns, and the cases of $\varepsilon_{xx}^{0} = 3\%$ and $\varepsilon_{xx}^{0} = 4\%$, which are run for 800 ps.

All MD simulations are performed with the Large-scale Atomic/Molecular Massively Parallel Simulator (LAMMPS) [48], using an integration timestep of 5 fs. Atomic configurations of the entire computational system, together with atomic-level stresses and potential energies, are recorded every 10 ps for the entire atomic system and every 1 ps for the surface region for subsequent analysis.

## 3. Results and discussion

The analysis of SAW-induced defect generation requires first establishing how the initially harmonic waves evolve during propagation, how this evolution depends on the crystallographic orientation of the surface, and how it modifies the stresses generated in the near-surface region. We therefore first analyze the nonlinear evolution of SAWs for the two crystallographic configurations considered, with particular attention to wave sharpening and the possible formation of a shock front. This provides the basis for the subsequent analysis of dislocation generation at the shock front, evolution of the generated defect structures under repeated interactions with the propagating wave, and the associated development of atomic-scale surface roughness.

### *3.1. Nonlinear SAW sharpening and shock formation*

To characterize the evolution of the elastic energy, strain, and stress fields during SAW propagation, the corresponding two-dimensional (2D) maps are constructed during post-processing of the MD results. The strain tensor is evaluated using a 2D grid with cell dimensions of approximately $1\ \mathrm{nm} \times 9\ nm \times 1.8\ \mathrm{nm}$. Atomic displacements are calculated relative to the atomic positions prior to the introduction of the SAW and then averaged over all atoms within each cell. The strain components are determined from the relative displacements of adjacent cells. The stress tensor in each cell is obtained by averaging the atomic-level stresses [50] over all atoms within the cell. A linear-elastic estimate of the local elastic strain energy per atom is obtained as

$$W = \frac{V_c}{2N_c}(\sigma_{xx}\varepsilon_{xx} + 2\sigma_{xz}\varepsilon_{xz} + \sigma_{zz}\varepsilon_{zz}), \tag{1}$$

where $V_c$ is the cell volume and $N_c$ are the volume and number of atoms in the cell, respectively.

Figure 2 illustrates the evolution of the elastic fields during propagation of a SAW with an initial strain amplitude of $\varepsilon_{xx}^0 = 1\%$ on the $\mathrm{Pt}(111)$ surface. The elastic strain energy, normal stress $\sigma_{xx}$, and shear stress $\sigma_{xz}$ are presented as 2D maps at selected times in a viewing window moving along the *x* direction with the SAW velocity *c* determined from the analysis presented in Appendix A and listed in Table A1.

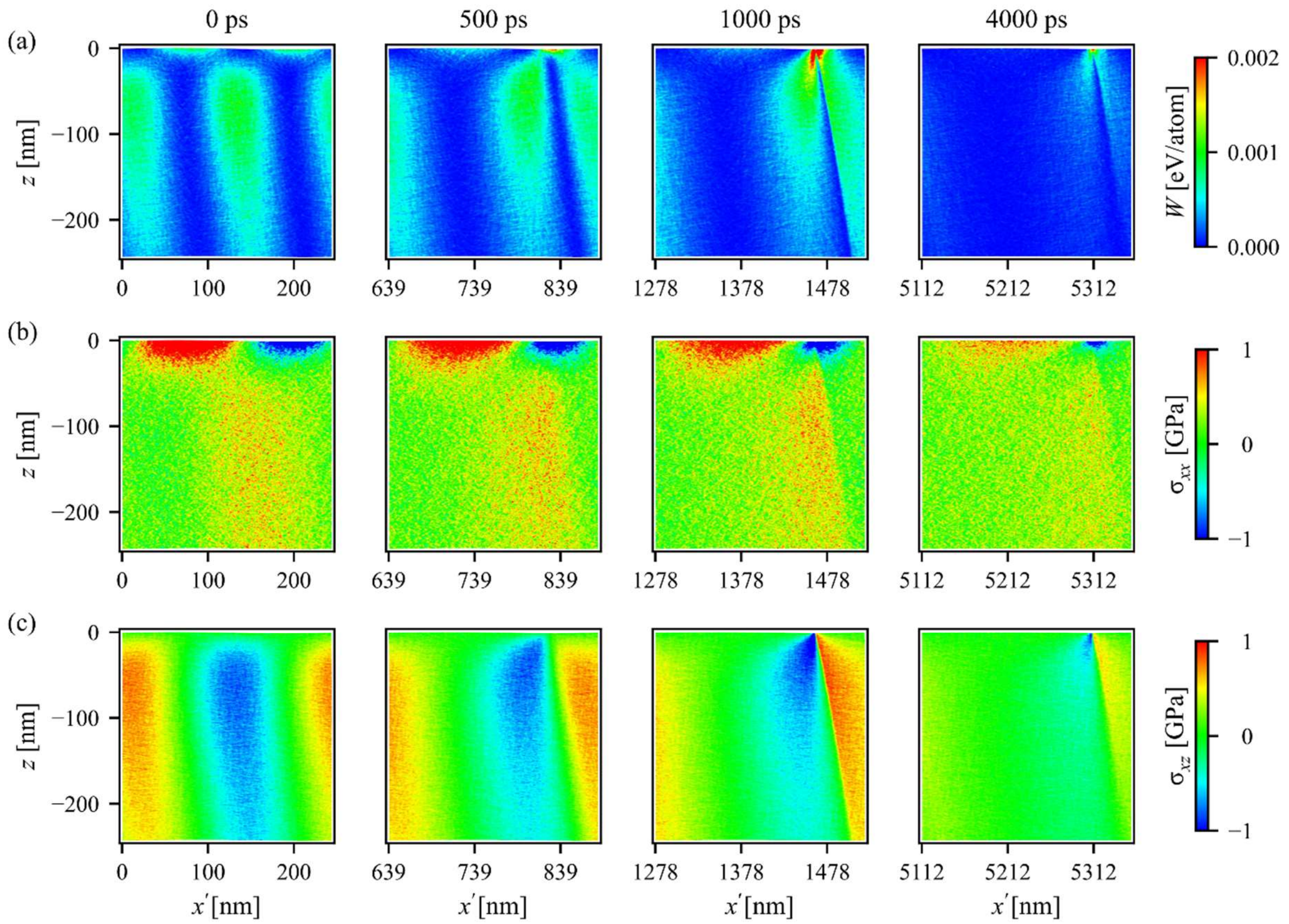

**Fig. 2.** Evolution of the elastic strain energy and stress fields in the near-surface region of the Pt(111) substrate during propagation of a SAW with an initial strain amplitude of $\varepsilon_{xx}^{0} = 1\%$. The figure presents 2D maps of (a) elastic strain energy $W$, (b) normal stress $\sigma_{xx}$ in the direction of wave propagation, and (c) shear stress $\sigma_{xz}$. The propagation time $t$ is indicated at the top of each panel, and the viewing window moves with the SAW. The horizontal coordinate is defined as $x' = x + ct$, where $c$ is the linear SAW velocity listed in Table A1. The maps illustrate nonlinear sharpening and shock front formation by about 1000 ps, followed by rapid dissipation of the SAW energy.

The maps provide a clear illustration of the major changes in the distribution of the elastic field that occur as the SAW propagates through the nonlinear medium. Starting from the strain and stress distributions predicted for a harmonic SAW within linear elasticity, the wave undergoes nonlinear transformation through the generation of higher-frequency harmonics [23,33,41-43,51], eventually leading to the formation of a sharp shock front. For a material with a positive parameter of local nonlinearity [33,52], corresponding to a positive quadratic correction to the linear stress–strain relation, the compressed portions of the wave propagate faster than the tensile portions, leading to progressive steepening of the wave profile and eventual shock formation. The resulting wave profile exhibits a characteristic asymmetry near the surface, as illustrated by the distributions of $\sigma_{xx}$ and $\sigma_{xz}$ in Fig. 2b,c.

The evolution of the wave profile shown in Fig. 2 also illustrates an important distinction between nonlinear SAWs and bulk acoustic waves. The nonlinear frequency up-conversion associated with SAW sharpening not only modifies the wave profile but also progressively concentrates the elastic strain energy near the surface, culminating in strong energy localization in the vicinity of the shock front, as illustrated by the map at 1000 ps in Fig. 2a. This distinctive behavior originates from the scaling of the characteristic penetration depth of a SAW with its wavelength [33,41,43]. The generation of higher-frequency harmonics during wave sharpening therefore redistributes the acoustic energy toward an increasingly thin surface layer. The resulting increase in the magnitudes of the normal and shear stresses (Fig. 2b,c) plays a key role in the generation of crystal defects discussed in Section 3.2.

At the initial strain amplitude of 1%, however, the shock front formation does not yet result in the generation of crystal defects. Instead, the onset of shock formation is accompanied by a rapid increase in the rate of wave energy dissipation, which is evident from the substantial reduction of the elastic strain energy by 4000 ps in Fig. 2a. While a detailed analysis of the mechanisms and channels of energy dissipation in nonlinear SAWs is presented elsewhere [53], we note that the dominant dissipation mechanism is activated only after the shock front has formed. At this stage, the high-frequency harmonics generated through nonlinear frequency up-conversion are efficiently converted into thermal phonons [33], resulting in rapid attenuation of the propagating wave. The absence of defect generation at $\varepsilon_{xx}^{0} = 1\%$ indicates that shock formation alone is insufficient to induce irreversible structural modification, although it fundamentally alters the dissipation of acoustic energy.

At higher initial strain amplitudes, nonlinear sharpening of the wave profile proceeds more rapidly and produces stronger localization of the elastic strain energy than in the $\varepsilon_{xx}^{0} = 1\%$ case discussed above. For $\varepsilon_{xx}^{0} = 4\%$, a well-defined shock front develops within about 200 ps, and the energy and stress maps in Fig. 3 reveal pronounced changes associated with the generation of subsurface dislocations. In particular, the elastic strain energy map at 400 ps (Fig. 3a) exhibits characteristic wave-scattering features and a horizontally aligned region of highly localized elastic energy at a depth of approximately 50 nm below the free surface. In the corresponding map of $\sigma_{xx}$ (Fig. 3b), this region is flanked by adjacent compressive and tensile stress fields, consistent with the presence of dislocations, as confirmed by the structural analysis presented in Section 3.2.

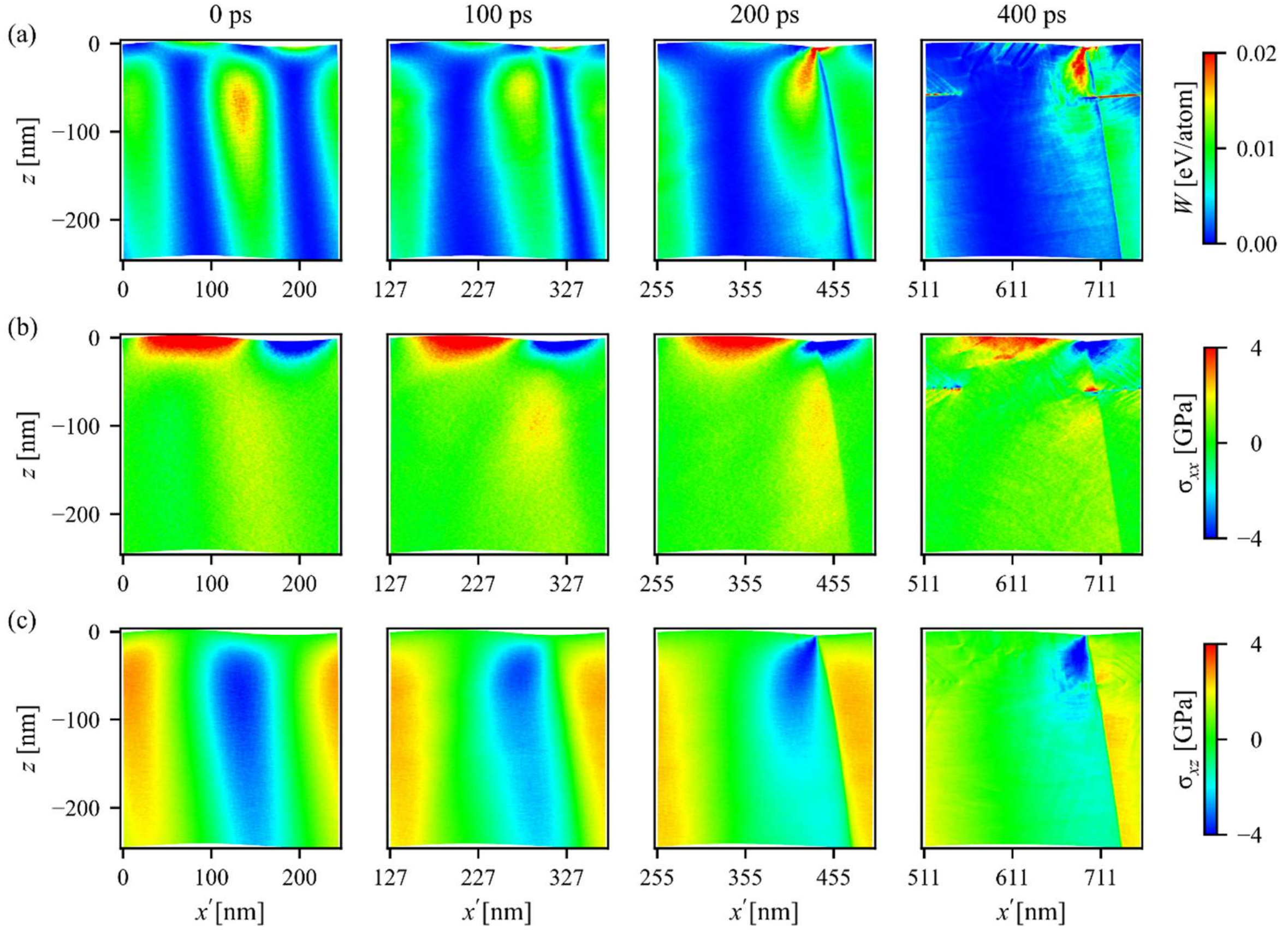

**Fig. 3.** Evolution of the elastic strain energy and stress fields in the near-surface region of the $\mathrm{Pt}(111)$ substrate during propagation of a SAW with an initial strain amplitude of $\varepsilon_{xx}^0 = 4\%$. The figure presents 2D maps of (a) elastic strain energy $W$, (b) normal stress $\sigma_{xx}$ in the direction of wave propagation, and (c) shear stress $\sigma_{xz}$. The propagation time $t$ is indicated at the top of each panel, and the viewing window moves with the SAW. The horizontal coordinate is defined as $x' = x + ct$, where $c$ is the linear SAW velocity listed in Table A1. Compared with the weaker SAW with $\varepsilon_{xx}^0 = 1\%$ shown in Fig. 2, nonlinear sharpening and shock front formation proceed more rapidly, with a well-defined shock front developing within about 200 ps. At 400 ps, pronounced wave-scattering features and localized regions of high stress appear in the subsurface layer, associated with interactions of the propagating wave with SAW-generated dislocations.

The generation of dislocations and subsequent interaction of the propagating wave with the resulting defect structures also provide an additional channel for dissipation of the SAW energy. Indeed, pronounced attenuation of the wave is already evident at 400 ps from the substantial reduction in the extent of regions with intermediate elastic strain energy (~0.01 eV) compared with the corresponding map at 200 ps, Fig. 3a. At the same time, the striped patterns in the instantaneous energy and stress distributions provide evidence of wave scattering by the generated defect structures.

After establishing the overall evolution of the elastic fields during SAW propagation, we turn to a quantitative analysis of nonlinear wave sharpening and shock front formation. In particular, the dependence of the shock formation distance, $L_{sh}$, on the initial strain amplitude $\varepsilon_{xx}^{0}$ can be used to determine the effective nonlinearity parameter $\beta$ for SAW propagation on a particular crystallographic surface and along a given propagation direction [33,54],

$$\beta = 1/L_{sh}k\varepsilon_{xx}^{0} \ , \quad (2)$$

where $k = 2\pi/\lambda$ is the wavenumber of the initially harmonic SAW.

Determination of $L_{sh}$ is complicated by the competition between nonlinear sharpening through frequency up-conversion and rapid dissipation of the high-frequency harmonics generated at the shock front. In the present study, the shock formation distance is determined from the evolution of the surface strain profile $\varepsilon_{xx}$ along the direction of wave propagation. The strain profile is evaluated at a depth of 1 nm below the surface to reduce the noise associated with calculations based on surface atoms. Since the values of $\varepsilon_{xx}$ calculated at the depth of 1 nm below the surface are within 4% of those evaluated for the surface atoms, in the discussion below we still refer to the result of the calculation as "surface strain" in the discussion below. Spatial averaging over grid cells with dimensions of approximately 1 nm × 9 nm × 0.7 nm reduces thermal fluctuations, while a discrete cosine transform (DCT) [55] is applied to further suppress thermal noise by filtering out high-frequency oscillations with wavelengths below 5 nm.

A series of strain profiles calculated for a SAW with an initial strain amplitude of $\varepsilon_{xx}^{0} = 1\%$ propagating on the Pt(111) surface is shown in Fig. 4. As the wave evolves, the compressive portion of the strain profile narrows and develops a pronounced cusp, while the tensile portion becomes progressively flattened. The compressive strain at the emerging cusp exceeds the initial strain amplitude by a factor of two at about 700 ps, reaches a maximum magnitude of approximately 2.9% at 1110 ps, and gradually decreases at later times due to wave dissipation. This asymmetric evolution of the wave profile is quantified by tracking the maximum and minimum values of $\varepsilon_{xx}$ during the first 2000 ps of wave propagation, as shown in Fig. 5a. The magnitude of the tensile strain decreases gradually and monotonically from the initial level of $\varepsilon_{xx}^0$ due to wave dissipation [53]. In contrast, the compressive strain increases in magnitude, particularly rapidly between 500 and 800 ps, before reaching a maximum as the shock front fully develops. The subsequent decrease reflects the increasing role of wave dissipation relative to nonlinear sharpening.

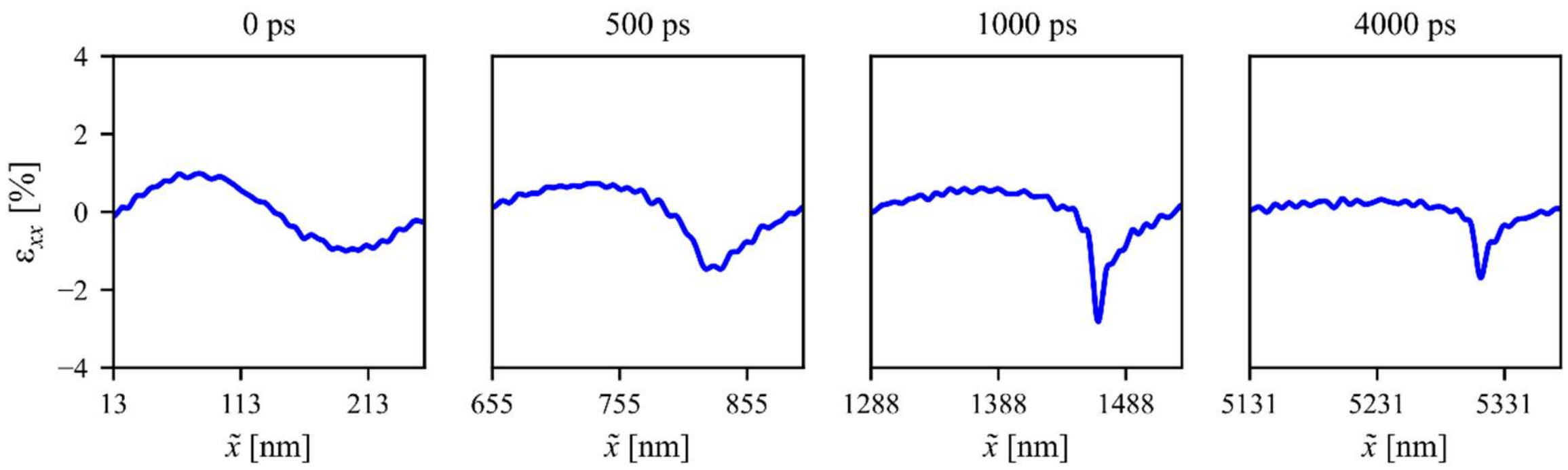

**Fig. 4.** Evolution of surface strain profile $\varepsilon_{xx}$ during propagation of a SAW with an initial strain amplitude of $\varepsilon_{xx}^0 = 1\%$ on the Pt(111) surface. To highlight the evolution of the waveform shape, each profile is cyclically shifted by $\Delta x_0(t)$, so that the strain is zero and has a positive local slope at the left boundary of the corresponding frame. The horizontal coordinate is then defined as $\tilde{x} = x' + \Delta x_0(t)$, where $x' = x + ct$ is the shifted coordinate used for the viewing window that follows the SAW at velocity $c$, as defined in Fig. 2. Progressive steepening of the compressive portion of the strain profile during the first 1000 ps reflects nonlinear frequency up-conversion and shock front formation. At later times, the wave amplitude decreases due to rapid dissipation of the SAW energy at the shock front.

To define the shock formation distance $L_{sh}$, we focus on the local curvature at the minimum of the compressive part of the strain profile, where the cusp develops during nonlinear wave sharpening. The local curvature is characterized by the quadratic coefficient $a_{\varepsilon_{xx}}$ obtained from a parabolic fit to the $\varepsilon_{xx}$ profile in the vicinity of the strain minimum, corresponding to the maximum compressive strain. The temporal evolution of $a_{\varepsilon_{xx}}$, plotted in Fig. 5b, exhibits a sigmoidal trend, with an initial slow increase followed by a nearly linear rapid-growth stage and a gradual approach to saturation. At later times, wave dissipation leads to a gradual decrease in the wave amplitude (Fig. 5a) and a corresponding slow reduction of the curvature at the compressive cusp (Fig. 5b). A sigmoid function is therefore fitted only over the time interval dominated by nonlinear sharpening, up to 1200 ps for the $\varepsilon_{xx}^{0} = 1\%$ case analyzed in Fig. 5. The shock formation time $\tau_{sh}$ is defined geometrically by the intersection of the tangent to the sigmoid at its inflection point with the horizontal line corresponding to the saturation level, as illustrated in Fig. 5b. The shock formation distance is then calculated as $L_{sh} = \tau_{sh}c$. For the SAW with $\varepsilon_{xx}^{0} = 1\%$ propagating on the Pt(111) surface, this procedure yields $\tau_{sh} = 771$ ps and $L_{sh} = 985.4$ nm. The same procedure is applied to SAWs with initial strain amplitudes of 0.5%, 1.5%, 2%, and 2.5% to determine the corresponding shock formation times and distances.

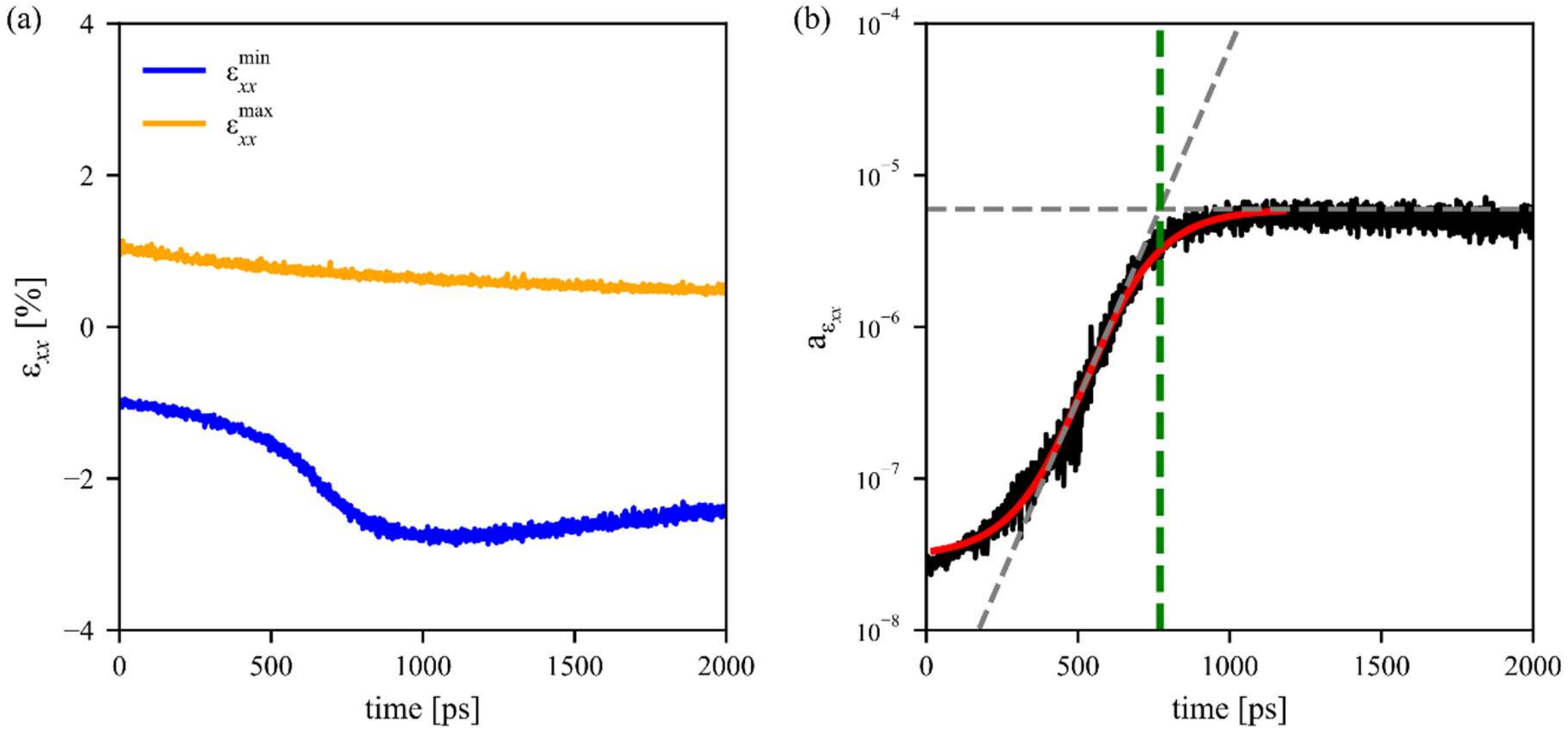

**Fig. 5.** Quantitative characterization of the evolution of surface strain profiles for a SAW with an initial strain amplitude of $\varepsilon_{xx}^{0} = 1\%$ propagating on Pt(111) surface. (a) Evolution of the maximum and minimum values of surface strain $\varepsilon_{xx}$. (b) Local curvature $a_{\varepsilon_{xx}}$ obtained from a parabolic fit near the strain minimum, where a cusp develops during shock front formation. The red curve in (b) shows a sigmoid fit to the curvature data up to 1200 ps. The gray dashed lines indicate the tangent at the midpoint of the sigmoid and its saturated value; their intersection defines the shock formation time $\tau_{sh}$, marked by the green vertical dashed line. The results are shown for the first 2000 ps of wave evolution.

An increase in the initial strain amplitude above $\varepsilon_{xx}^0 = 2.5\%$ produces substantial changes in the evolution of the wave profile, as illustrated in Figs. 6 and 7 for $\varepsilon_{xx}^0 = 4\%$. The initial evolution remains similar to that discussed above for the weaker wave, with rapid nonlinear sharpening and development of a sharp cusp on the compressive side of the strain profile. The compressive strain increases from 4% to approximately 7.0% within only 180 ps, illustrating the much faster nonlinear evolution compared with the $\varepsilon_{xx}^0 = 1\%$ case. Beyond 180 ps, however, oscillations appear in the strain profile, becoming clearly visible at 200 ps and more pronounced by 400 ps (Fig. 6). Correspondingly, both the strain extrema (Fig. 7a) and the curvature at the compressive minimum (Fig. 7b) exhibit pronounced irregular oscillations after approximately 180 ps. Similar to the wave-scattering features observed for the same simulation in Fig. 3, the appearance of these oscillations can be attributed to the emission of dislocations, as discussed in detail in Section 3.2. Despite these oscillations, the overall sigmoidal trend of $a_{\varepsilon_{xx}}$ remains discernible in Fig. 7b, making it possible to extract $L_{sh}$ from a sigmoid fit after convolution-based smoothing of the data. The shock formation time obtained for $\varepsilon_{xx}^0 = 4\%$ is very close to 200 ps, corresponding to a shock formation distance comparable to the wavelength λ of the fundamental harmonic of the initial wave.

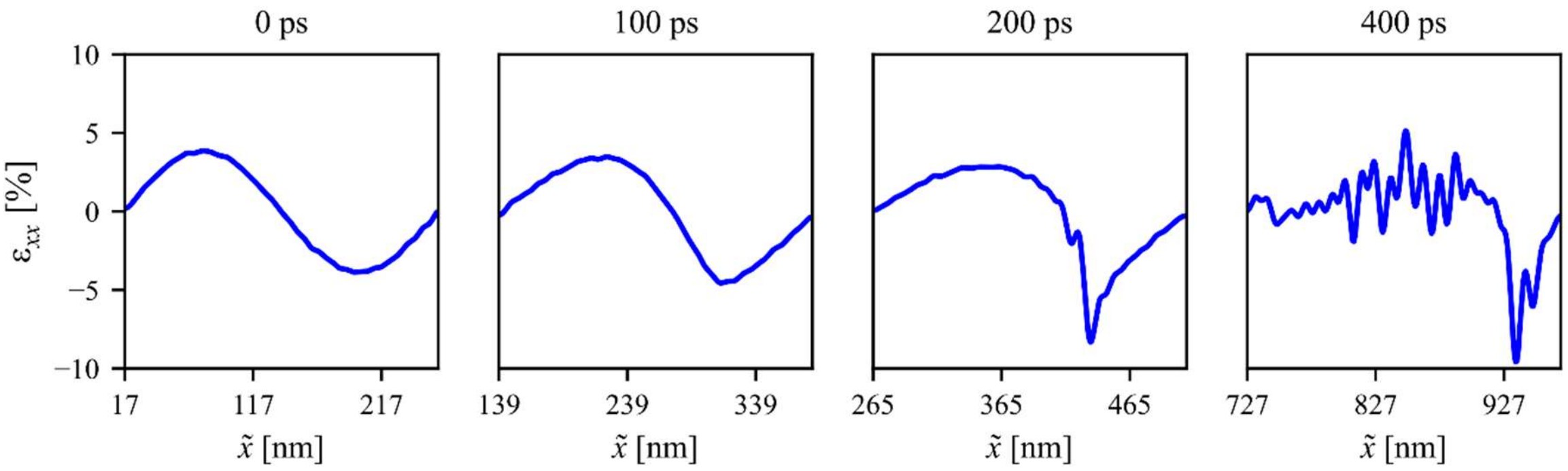

**Fig. 6.** Evolution of surface strain profile $\varepsilon_{xx}$ during propagation of a SAW with an initial strain amplitude of $\varepsilon_{xx}^0 = 4\%$ on the Pt(111) surface. The coordinate $\tilde{x}$ used for the $x$-axis is defined as described in the caption of Fig. 4. Nonlinear sharpening proceeds more rapidly than for $\varepsilon_{xx}^0 = 1\%$ (Fig. 4), and the oscillations apparent in the profiles at 200 and 400 ps are associated with SAW-induced dislocation generation.

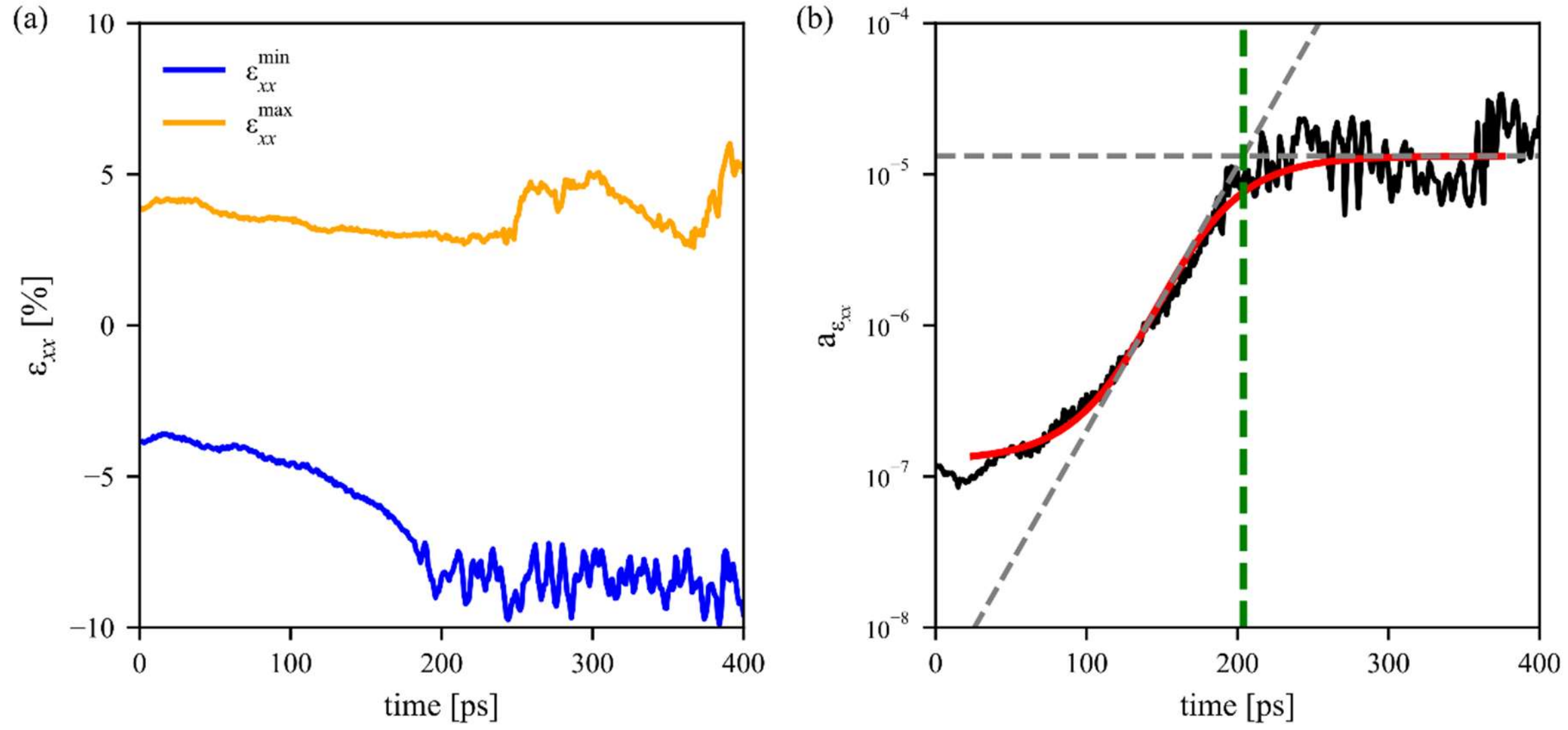

**Fig. 7.** Quantitative characterization of the evolution of surface strain profiles for a SAW with an initial strain amplitude of $\varepsilon_{xx}^0 = 4\%$ propagating on $\mathrm{Pt}(111)$ surface. (a) Evolution of the maximum and minimum values of surface strain $\varepsilon_{xx}$. (b) Local curvature $a_{\varepsilon_{xx}}$ obtained from a parabolic fit near the strain minimum, where a cusp develops during shock front formation. The red curve in (b) shows a sigmoid fit to the curvature data over the time interval from 25 to 375 ps. The gray dashed lines indicate the tangent at the midpoint of the sigmoid and its saturated value; their intersection defines the shock formation time $\tau_{sh}$, marked by the green vertical dashed line. Oscillations appearing after approximately 180 ps are associated with SAW-induced dislocation generation.

The effective nonlinearity parameter $\beta$ is determined from the shock formation distances obtained in simulations below the threshold for dislocation generation ($\varepsilon_{xx}^0$ = 0.5%, 1%, 1.5%, 2%, and 2.5%). According to Eq. (2), $L_{sh}$ is expected to scale linearly with the reciprocal of the initial strain amplitude, $1/\varepsilon_{xx}^0$. A linear fit to this dependence, shown in Fig. 8, yields a coefficient of determination $R^2 = 0.99331$, demonstrating excellent agreement with the expected scaling. The data points for $\varepsilon_{xx}^0$ = 3% and 4% are excluded from the fit to avoid any influence of dislocation emission on the determination of $\beta$. Interestingly, these points nevertheless closely follow the same linear trend established by the lower-amplitude simulations. This observation suggests that dislocation emission occurs close to the time of shock formation, so that the associated distortions of the strain profiles do not substantially affect the values of $L_{sh}$ determined by the procedure described above. The resulting value of $\beta$ = 3.864 for a SAW propagating along the $[11\bar{2}]$ direction on the $\mathrm{Pt}(111)$ surface falls within the range of values reported for bulk longitudinal waves [56,57] and SAWs [54] in metals.

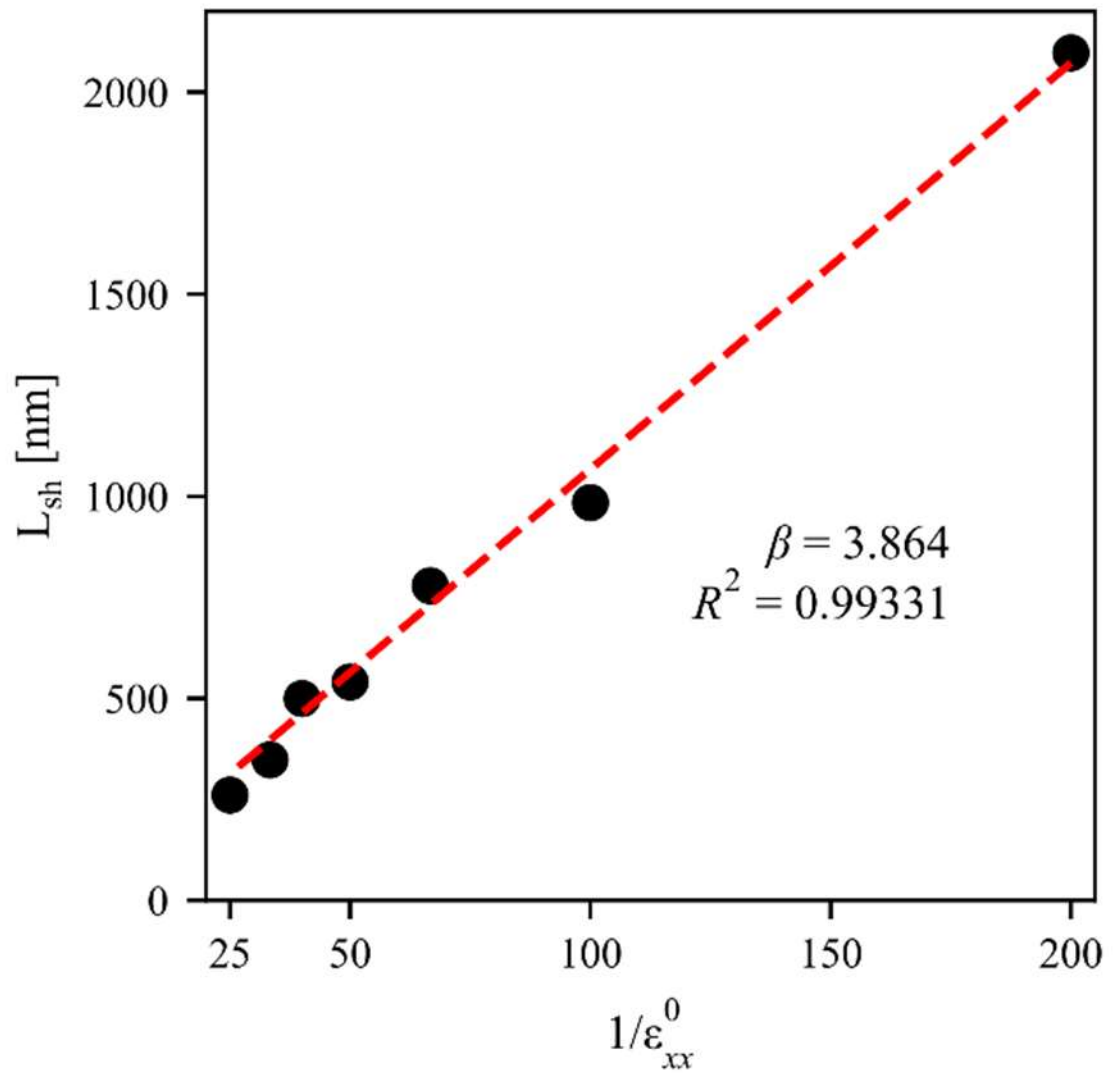

**Fig. 8.** Shock formation distance $L_{sh}$ as a function of the reciprocal initial strain amplitude $1/\varepsilon^0_{xx}$ for SAWs propagating along the $[11\bar{2}]$ direction on the $\mathrm{Pt}(111)$ surface. The linear fit is based on simulations with $\varepsilon^0_{xx}$ = 0.5–2.5%, in which no dislocation emission is observed. The data points for $\varepsilon^0_{xx}$ = 3% and 4%, where dislocation emission occurs, are excluded from the fit but remain close to the same linear dependence. The slope of the fitted line is used to determine the effective nonlinearity parameter $\beta$ according to Eq. (2).

SAW propagation along the $[100]$ direction on the $\mathrm{Pt}(001)$ surface exhibits markedly different behavior from that discussed above for $\mathrm{Pt}(111)$. As illustrated in Fig. 9 for an initial strain amplitude of $\varepsilon^0_{xx}$ = 2%, the surface strain profile shows little nonlinear evolution over the 2000 ps simulation. Its nearly sinusoidal shape is largely preserved, with only a slight reduction in amplitude and minor steepening of the ascending branch of the strain profile.

The strong dependence of nonlinear SAW evolution on surface orientation and propagation direction is further demonstrated in Fig. 10, where the temporal evolution of the strain extrema and local curvature is compared for the $(111)[11\bar{2}]$ and $(001)[100]$ SAWs with the same initial strain amplitude of $\varepsilon^0_{xx}$ = 2%. For the $(001)[100]$ SAW, the magnitudes of both strain extrema exhibit only a small decrease over the simulation time (Fig. 10c), in sharp contrast to the transient increase in compressive strain followed by rapid attenuation observed for the $(111)[11\bar{2}]$ SAW (Fig. 10a). Similarly, the curvature coefficient $a_{\varepsilon_{xx}}$ at the strain minimum remains nearly constant for the $(001)[100]$ SAW (Fig. 10d), whereas a pronounced sigmoidal increase associated with nonlinear sharpening and shock formation is observed for the $(111)[11\bar{2}]$ SAW (Fig. 10b). The much weaker attenuation of the $(001)[100]$ SAW also highlights the important role of nonlinear sharpening and shock front formation in accelerating SAW energy dissipation [53].

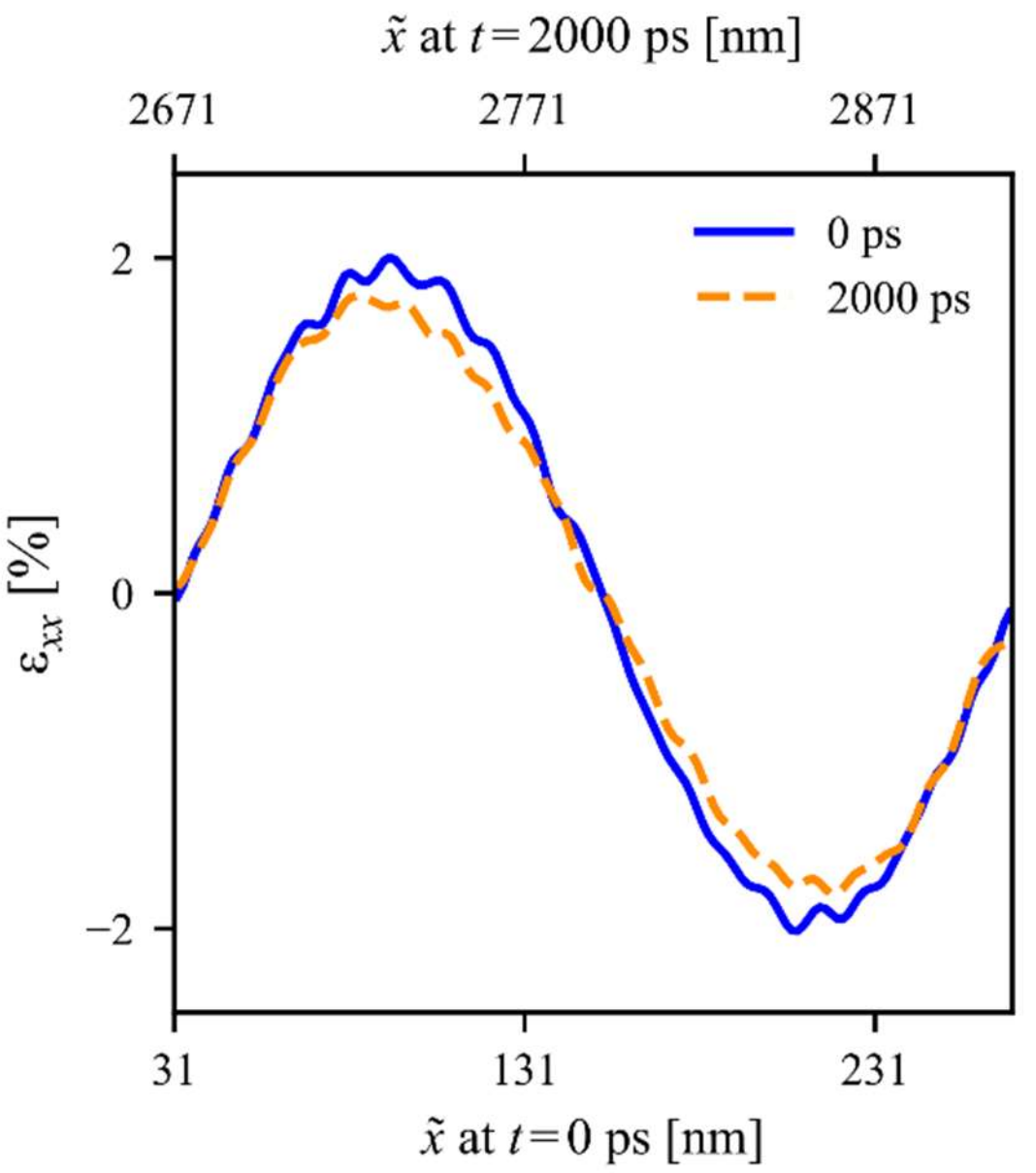

**Fig. 9.** Evolution of the surface strain profile $\varepsilon_{xx}$ during propagation of a SAW with an initial strain amplitude of $\varepsilon_{xx}^0 = 2\%$ along the $[100]$ direction on the $\mathrm{Pt}(001)$ surface, illustrated by profiles at 0 ps and 2000 ps. The coordinate $\tilde{x}$ used for the $x$-axis is defined as described in the caption of Fig. 4. The waveform retains its nearly sinusoidal shape over the simulated propagation time, in contrast to the pronounced nonlinear sharpening observed for SAWs on Pt(111) (Figs. 4 and 6).

The results in Figs. 9 and 10 suggest that nonlinear frequency up-conversion for the $(001)[100]$ SAW proceeds too slowly to produce a sharp shock front on the timescale of the simulations. Indeed, the gradual decrease in wave amplitude (Fig. 10c) is more pronounced than the barely discernible increase in local curvature (Fig. 10d), indicating that dissipation competes effectively with the weak nonlinear sharpening. These trends suggest that dissipation is likely to overtake nonlinear sharpening at longer propagation times, preventing the development of a well-defined shock front. The absence of appreciable nonlinear sharpening persists in simulations with initial strain amplitudes up to $\varepsilon_{xx}^0$ = 6%. At $\varepsilon_{xx}^0$= 8%, nonlinear waveform sharpening becomes discernible, with a shock formation time of 1543 ps determined from the evolution of the local curvature (Fig. S1 in the Supplementary Material). The associated increase in compressive strain, however, is offset by pronounced dissipation of the wave energy, as reflected in the gradual decrease in the magnitude of the compressive strain from about 200 ps until the end of the simulation at 2800 ps. Thus, although nonlinear sharpening occurs at this high initial amplitude, its effect on the strain profile is masked by the dominant wave dissipation.

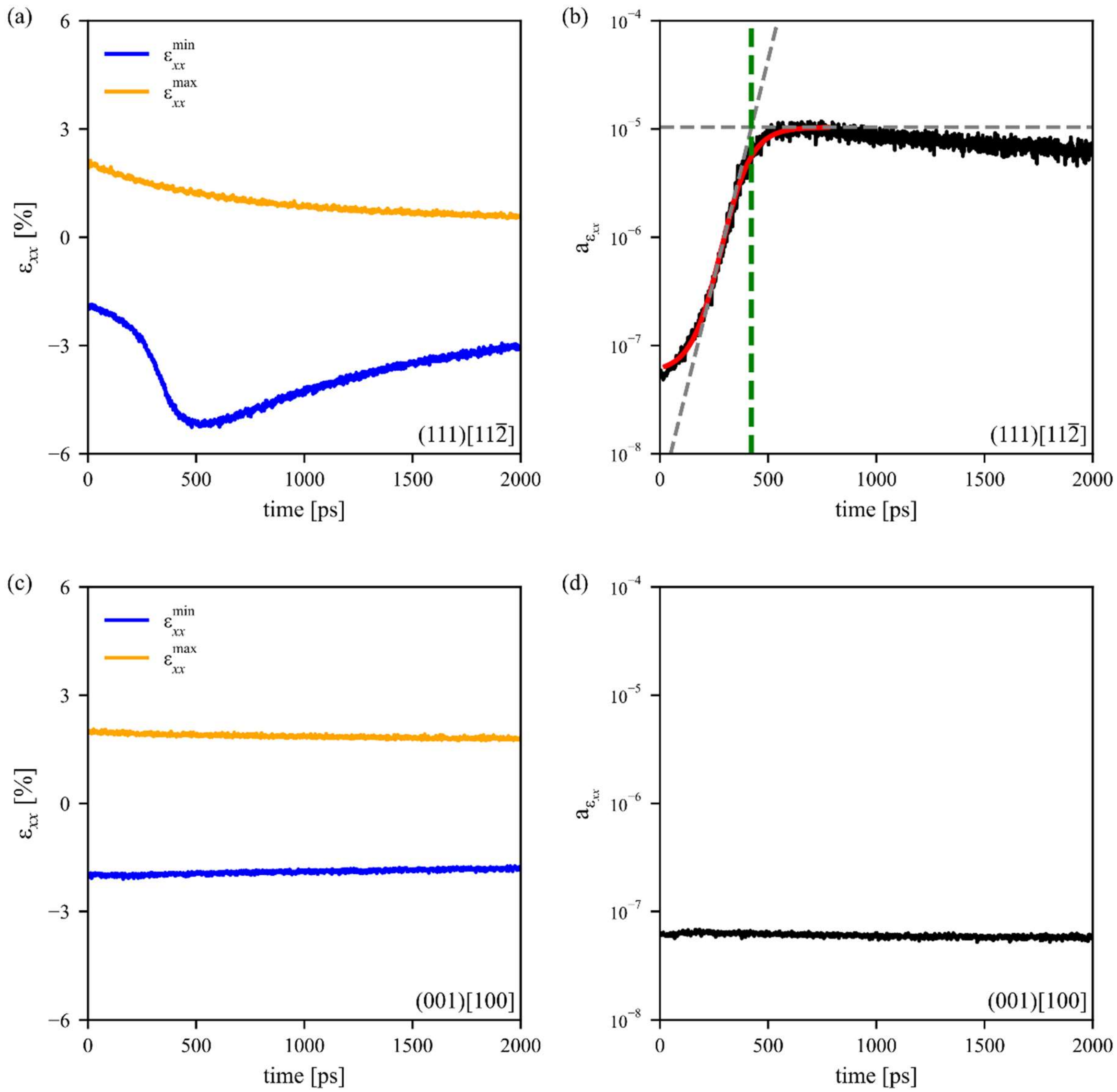

**Fig. 10.** Comparison of the evolution of surface strain profiles for SAWs with initial strain amplitudes of $\varepsilon_{xx}^{0} = 2\%$ propagating along the $[11\bar{2}]$ direction on the $(111)$ surface (a,b) and along the $[100]$ direction on the $(001)$ surface (c,d). For the $(111)[11\bar{2}]$ SAW, nonlinear sharpening leads to a transient increase in the magnitude of compressive strain (a) and a pronounced increase in the local curvature at the strain minimum (b). In contrast, for the $(001)[100]$ SAW, no significant changes are observed in either the strain amplitude (c) or the local curvature (d) over the simulated propagation time.

The striking differences between the two cases considered above highlight the strong dependence of nonlinear SAW evolution not only on the nonlinear elastic moduli governed by the crystal structure and anharmonicity of the interatomic interactions, but also on the crystallographic surface orientation and direction of wave propagation [43,58]. The weak nonlinear sharpening observed for the $(001)[100]$ SAW in Pt is not, however, a general feature of this surface orientation and propagation direction. Indeed, earlier MD simulations of $(001)[100]$ SAW propagation in an fcc solid described by the Lennard–Jones (LJ) potential revealed pronounced nonlinear sharpening and formation of a sharp shock front for initial strain amplitudes ranging from 2% to 4.7%, with an effective nonlinearity parameter $\beta$ = 0.85 determined from Eq. (2) [33].

To explore the origin of this difference, the nonlinear elastic response predicted by the EAM potential used in the present study is compared with that of an LJ potential parametrized to reproduce the lattice constant and $C_{11}$ elastic modulus of Pt. Bulk fcc crystals described by the two potentials are subjected to uniaxial strain along the [100] direction, and the resulting stress–strain dependences $\sigma_{xx}(\varepsilon_{xx})$ are shown in Fig. S2 of the Supplementary Material. The comparison indicates stronger stiffening under compression and softening under tension for the LJ potential. The weaker physical nonlinearity predicted by the EAM potential, combined with the opposite sign of the geometric nonlinearity for an fcc crystal along the [100] direction [33], may therefore lead to partial compensation of the two contributions and provide a qualitative explanation for the weak nonlinear evolution of the $(001)[100]$ SAWs observed in the present simulations. This interpretation should be regarded as qualitative, however, since nonlinear SAW evolution is inherently nonlocal and involves coupling among multiple strain components. The effective SAW nonlinearity therefore cannot be inferred quantitatively from a uniaxial stress–strain response alone.

### *3.2. SAW-induced generation and evolution of dislocation structures*

The nonlinear sharpening of the SAW profile and shock front formation discussed in Section 3.1 produce highly localized transient stresses in the near-surface region that can initiate plastic deformation. In this section, we examine the conditions for dislocation emission at the SAW shock front and follow the subsequent evolution of the generated dislocation structures through interactions and reactions driven by the propagating wave.

#### *3.2.1. Onset of dislocation emission near the theoretical shear strength*

To identify structural changes produced in the Pt substrate by SAWs undergoing nonlinear transformation, atomic configurations are recorded at 1 ps intervals and quenched at an extremely high cooling rate of approximately $6\times10^{14}$ K/s to reduce thermal noise. Dislocation analysis is then performed for the quenched configurations using the Dislocation Extraction Algorithm (DXA) implemented in OVITO [59,60].

For the SAW with an initial strain amplitude of $\varepsilon_{xx}^{0} = 4\%$ propagating on the Pt(111) surface, no structural changes in the substrate are detected during the first 180 ps of wave propagation. At 180 ps, very close to the shock formation time $\tau_{sh}$ identified from the analysis illustrated in Fig. 7b, surface nucleation of the first dislocation half-loop is observed, as shown in Fig. 11a. The dislocation is identified as a Shockley partial with a Burgers vector of $\frac{1}{6}[112]$ gliding on the $(11\bar{1})$ plane oriented with respect to the SAW propagation direction as illustrated in Fig. 11d.

The delayed dislocation emission indicates that the local stress state near the substrate surface reaches the critical condition for slip activation only after substantial nonlinear sharpening of the SAW profile. Since the resolved shear stress (RSS), $\tau_{\mathrm{RSS}}$, on a potential slip system provides a natural measure of the local driving force for dislocation nucleation and glide, the evolution of the RSS is analyzed for all twelve $\{111\}\langle112\rangle$ Shockley partial slip systems by projecting the local stress tensor onto the corresponding slip planes and directions. To evaluate the conditions for dislocation emission from the surface, the RSS is first analyzed within the same 0.7-nm-thick layer located 1 nm below the surface that is used to obtain the one-dimensional strain profiles in Figs. 4-7 and 9-10. Similar to the strain profiles, a DCT-based filter is applied to the RSS profiles to remove high-frequency thermal fluctuations with wavelengths below 5 nm.

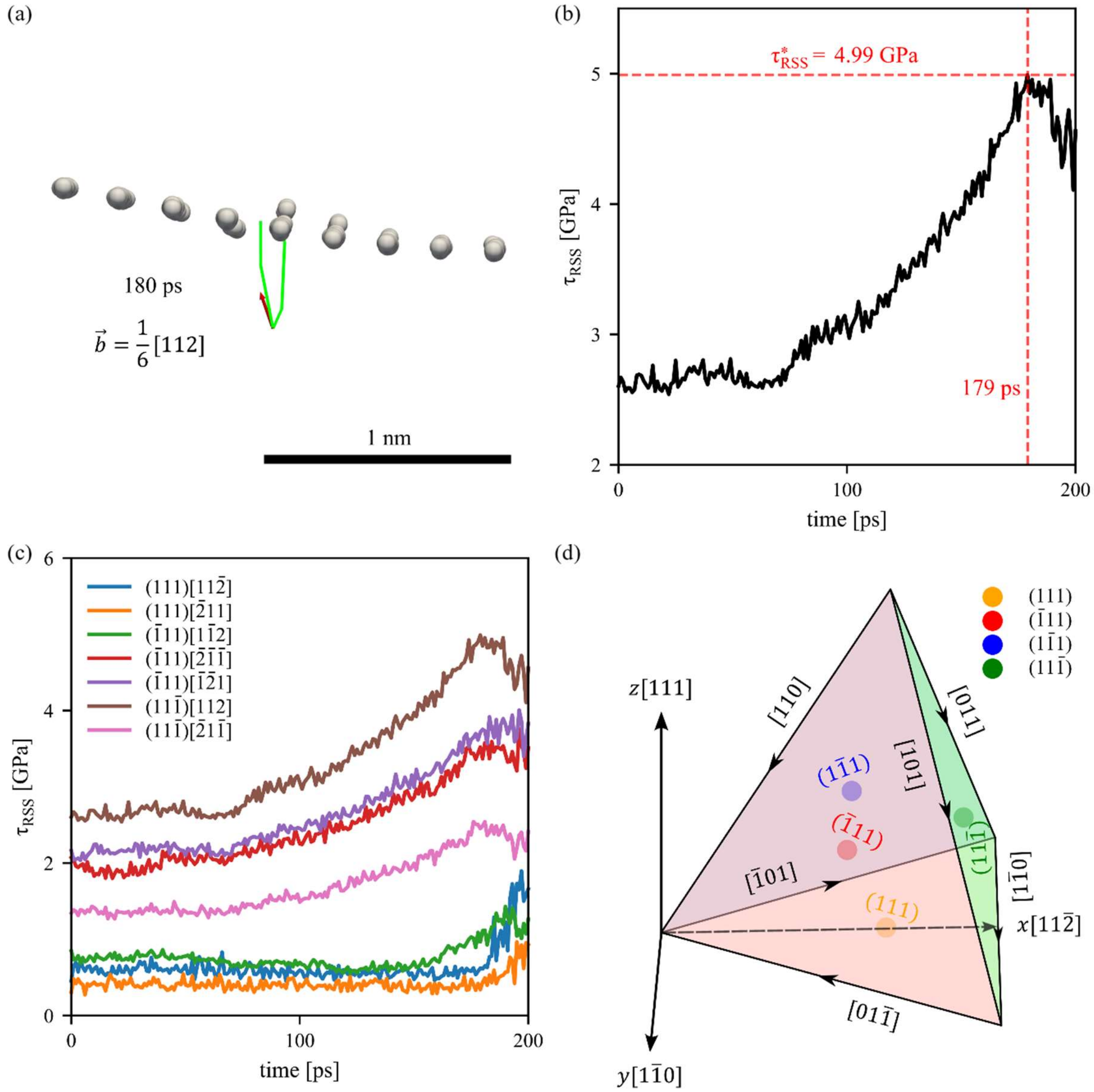

**Fig. 11.** Emission of the first dislocation and analysis of the RSS in a simulation of a SAW with an initial strain amplitude of $\varepsilon^0_{xx} = 4\%$ propagating on the Pt$(111)$ surface. (a) Snapshot at the moment of surface nucleation of the first Shockley partial dislocation half-loop. The red arrow indicates the Burgers vector, and the gray spheres represent surface atoms. (b) Evolution of the peak value of near-surface RSS for the $(11\bar{1})[112]$ slip system, leading to the emission of the first dislocation at 180 ps. The maximum level of RSS observed in the simulation, $\tau^{*}_{\mathrm{RSS}}$ = 4.99 GPa, and the time when this level is reached are marked by the horizontal and vertical red dashed lines, respectively. (c) Comparison of the time evolution of peak near-surface RSS values for all Shockley partial slip systems during the first 200 ps of the simulation, with only one representative

system shown for each symmetry-equivalent set. (d) Schematic illustration of the $\{111\}\langle 112\rangle$ Shockley partial slip systems in an fcc crystal. The Pt(111) surface is normal to the *z* axis, and the SAW propagates along the *x* axis.

An example of the evolution of the RSS profile is shown in Fig. 12 for the $(11\bar{1})[112]$ slip system on which the first dislocation half-loop is emitted. As the SAW undergoes nonlinear sharpening, the peak RSS increases and reaches its maximum at 179 ps, immediately before the first dislocation half-loop is detected in the configuration recorded at 180 ps. The peak RSS then gradually decreases. The evolution of the peak RSS for this slip system is plotted in Fig. 11b, where its maximum value, $\tau^{*}_{\mathrm{RSS}}$ = 4.99 GPa, is marked by the horizontal dashed line. This value is remarkable because it is close to the theoretical shear strength of a perfect crystal. The classical Frenkel estimate gives a critical shear stress of the order of $\tau_c \approx G/10$, where *G* is the appropriate shear modulus [61]. For the $\{111\}\langle 112\rangle$ shear deformation considered here, the relaxed and unrelaxed shear moduli are given by $G_r^{111} = 3C_{44}(C_{11} - C_{12})/(C_{11} - C_{12} + 4C_{44})$ and $G_u^{111} = (C_{11} - C_{12} + C_{44})/3$, respectively [62]. The relaxed modulus corresponds to deformation with all stress components other than the principal shear stress relaxed to zero, whereas the unrelaxed modulus corresponds to deformation with all strain components other than the principal shear strain constrained to zero. Using the room-temperature elastic constants predicted by the EAM Pt potential (Section 2) gives $G_r^{111} = 29.1$ GPa and $G_u^{111} = 38.3$ GPa. Thus, the maximum RSS associated with the onset of SAW-induced dislocation emission, $\tau^{*}_{\mathrm{RSS}}$ = 4.99 GPa, corresponds to $0.17G_r^{111}$ and $0.13G_u^{111}$.

Although the Frenkel approximation is generally considered to overestimate the theoretical shear strength [63], a number of *ab initio* calculations predict ideal shear strengths between 8% and 15% of the corresponding shear moduli for relaxed and unrelaxed shear deformation [62,64-66]. Several characteristics of the SAW-induced deformation may enable the local shear stress to approach the theoretical shear strength of a perfect crystal. These include the ultrahigh loading rate at the SAW shock front, the rapid increase in shear stress during the final stage of shock front formation (Fig. 11b), and the complex transient stress state generated by the rapidly advancing shock front. The importance of the latter is supported by computational studies demonstrating a strong sensitivity of the ideal shear strength to the local multiaxial stress state, including stresses acting on the slip plane in directions other than the slip direction [66,67]. The stresses approaching the theoretical shear strength are particularly notable given that dislocation nucleation at the SAW shock front occurs at 300 K in the vicinity of a free surface, where thermal fluctuations and the presence of the surface may lower the stress required for nucleation relative to the ideal strength of a perfect bulk crystal.

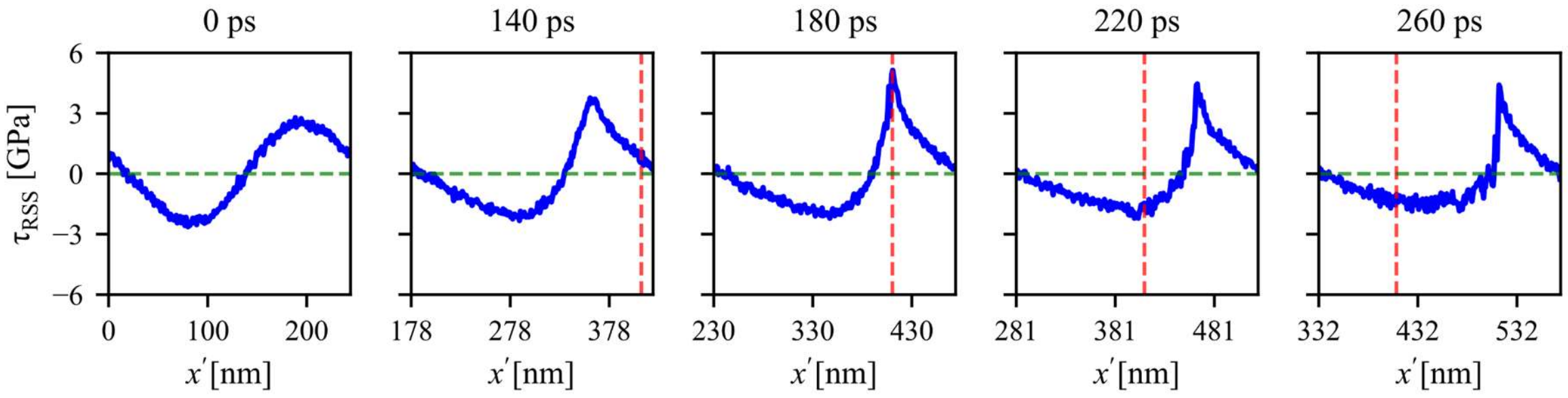

**Fig. 12.** Evolution of the RSS profile for the $(11\bar{1})[112]$ slip system during propagation of a SAW with an initial strain amplitude of $\varepsilon_{xx}^{0} = 4\%$ on the Pt(111) surface. The RSS is calculated in the near-surface region, 1 nm below the surface. The propagation time $t$ is indicated at the top of each panel, and the viewing window is moving with the SAW. The horizontal coordinate is shifted according to $x' = x + ct$, where $c$ is the linear SAW velocity listed in Table A1. The vertical red dashed line marks the location at which emission of the first dislocation half-loop is observed. The time dependence of the peak value of $\tau_{\text{RSS}}$ is shown in Fig. 11b.

The prediction that the shear stress at the onset of dislocation emission approaches the theoretical shear strength suggests that the generation and probing of strong nonlinear SAWs may provide a route for direct measurement of the ideal shear strength of crystals with a low density of preexisting defects in the near-surface region. Direct experimental access to ideal shear strength has long been challenging because defect-mediated plasticity generally intervenes before the theoretical limit is reached [68]. Nonlinear SAWs could therefore complement existing approaches that suppress or delay heterogeneous defect nucleation by reducing sample dimensions, as in whiskers and micro- or nanopillars [69-71], and/or confining the highly stressed volume through nanoindentation [72,73].

### *3.2.2. Activation of additional slip systems and evolution of dislocation structures*

Following the emission of the first dislocation on the $(11\bar{1})[112]$ slip system, for which the highest RSS is reached, additional dislocations progressively nucleate as the shock front propagates along the surface. This continuous emission of dislocations at the shock front is illustrated in Fig. 13, where a sequence of snapshots shows the expansion of the subsurface region affected by dislocation activity. Dislocation half-loops are emitted not only on the $(11\bar{1})[112]$ slip system but also on four additional systems on the $(1\bar{1}1)$ and $(\bar{1}11)$ slip planes, for which the near-surface RSS reaches levels comparable to that on the $(11\bar{1})[112]$ system (Fig. 11c). Because the half-loops nucleate in close proximity to one another, Shockley partials gliding on intersecting planes can react to form $\frac{1}{6}\langle 110\rangle$ stair-rod dislocation segments along the intersections of the slip planes. The sessile character of these segments temporarily pins the resulting dislocation structures behind the shock front, as illustrated by the snapshot at 190 ps in Fig. 13.

Moreover, the rapid spatial and temporal variations of the stress field in the near-surface region around the SAW shock front, illustrated by the RSS profiles for the $(11\bar{1})[112]$ slip system in Fig. 12, produce changes in the preferred slip planes of the expanding Shockley partial half-loops and promote multiple cross-slip events between close-packed planes. Each such event produces a stair-rod segment through a dislocation reaction, such as $\frac{1}{6}[112] \rightarrow \frac{1}{6}[0\bar{1}1] + \frac{1}{6}[121]$ for cross-slip from $(11\bar{1})$ to $(1\bar{1}1)$ slip plane. As in the reactions between intersecting Shockley partials discussed above, the resulting $\frac{1}{6}\langle 110\rangle$ stair-rod segments are sessile and present barriers to further dislocation motion.

An example of dislocation reactions involving such cross-slip events is shown in Fig. 14. Shockley partial half-loops initially emitted on the $(11\bar{1})$ slip plane undergo multiple cross-slip events to the $(1\bar{1}1)$ and $(\bar{1}11)$ slip planes within only several picoseconds. The configurations, plotted at 1 ps intervals, reveal substantial changes in the dislocation configuration between consecutive frames, demonstrating the extremely rapid character of these reactions. As a result, the dislocation configuration rapidly transforms into the surface-anchored "swing-like" structure marked **A** in the snapshot at 200 ps in Fig. 13. This structure consists of a $\frac{1}{2}[110]$ dislocation anchored to the surface, with its segments on the $(\bar{1}11)$ to $(1\bar{1}1)$ slip planes dissociated into pairs of Shockley partials separated by stacking faults and connected by a sessile $\frac{1}{6}[1\bar{1}0]$ segment at the intersection of the two planes. Similar structures form at multiple locations behind the SAW shock front, as can be seen in Fig. 13.

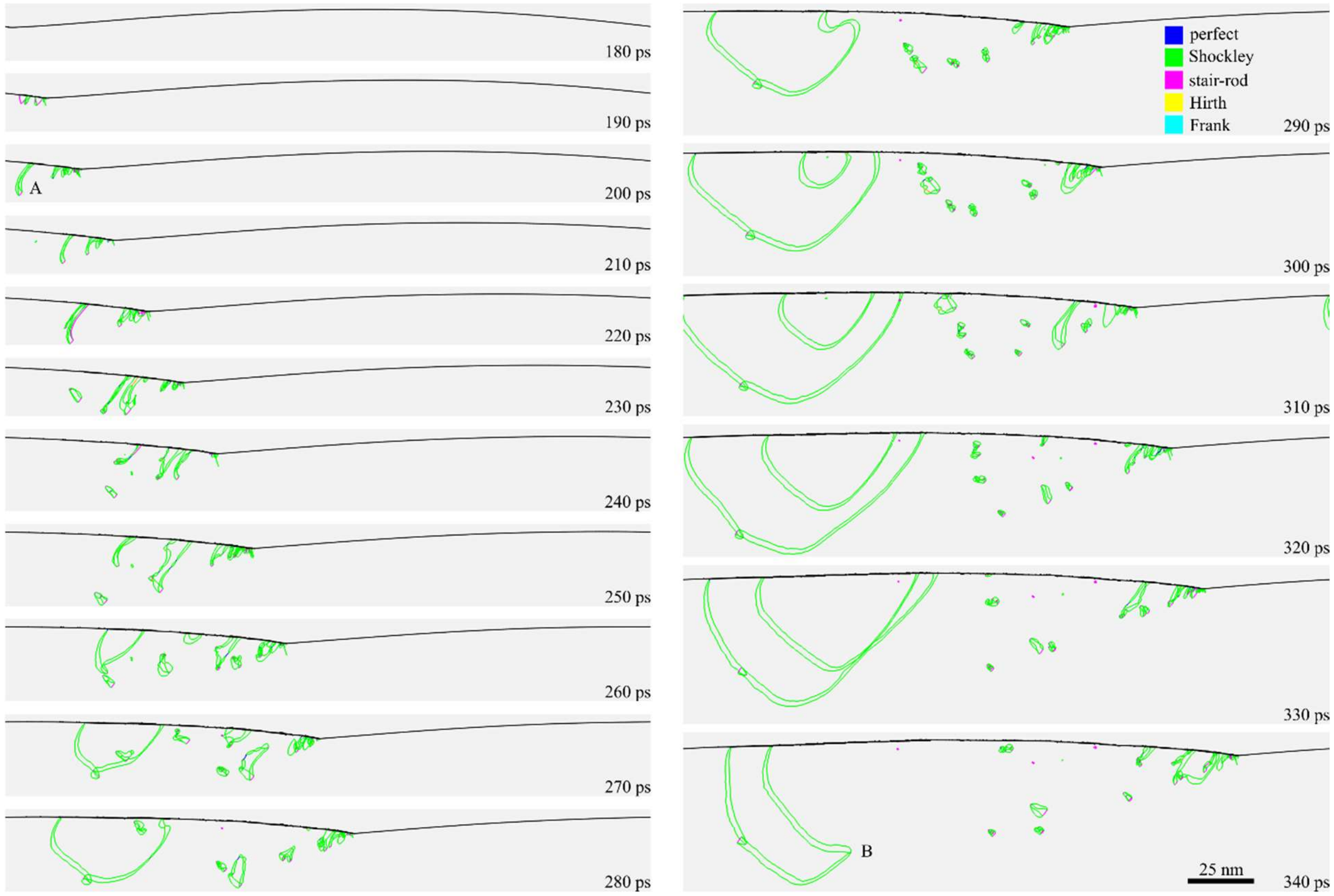

**Fig. 13.** Snapshots of the dislocation structures generated by a SAW with an initial strain amplitude of $\varepsilon_{xx}^{0} = 4\%$ on the Pt(111) surface. Dislocations nucleate at the SAW shock front and interact in the region behind the propagating wave. The displayed domain spans the full system length along the propagation direction, with its left boundary chosen to coincide with the location at which the first dislocation is observed. The displayed depth $\Delta z$ is adjusted to ensure that all dislocations are visible: $\Delta z$ = 25 nm for $t$ = 180–230 ps, $\Delta z$ = 30 nm for $t$ = 240–280 ps, $\Delta z$ = 50 nm for $t$ = 290–320 ps, and $\Delta z$ = 60 nm for $t$ = 330–340 ps. Dislocation types are color coded according to the legend shown in the panel at $t$ = 290 ps. Dislocation reactions leading to the formation of the half-loops marked **A** in the panel at $t$ = 200 ps and **B** in the panel at $t$ = 340 ps are detailed in Figs. 14 and 15, respectively.

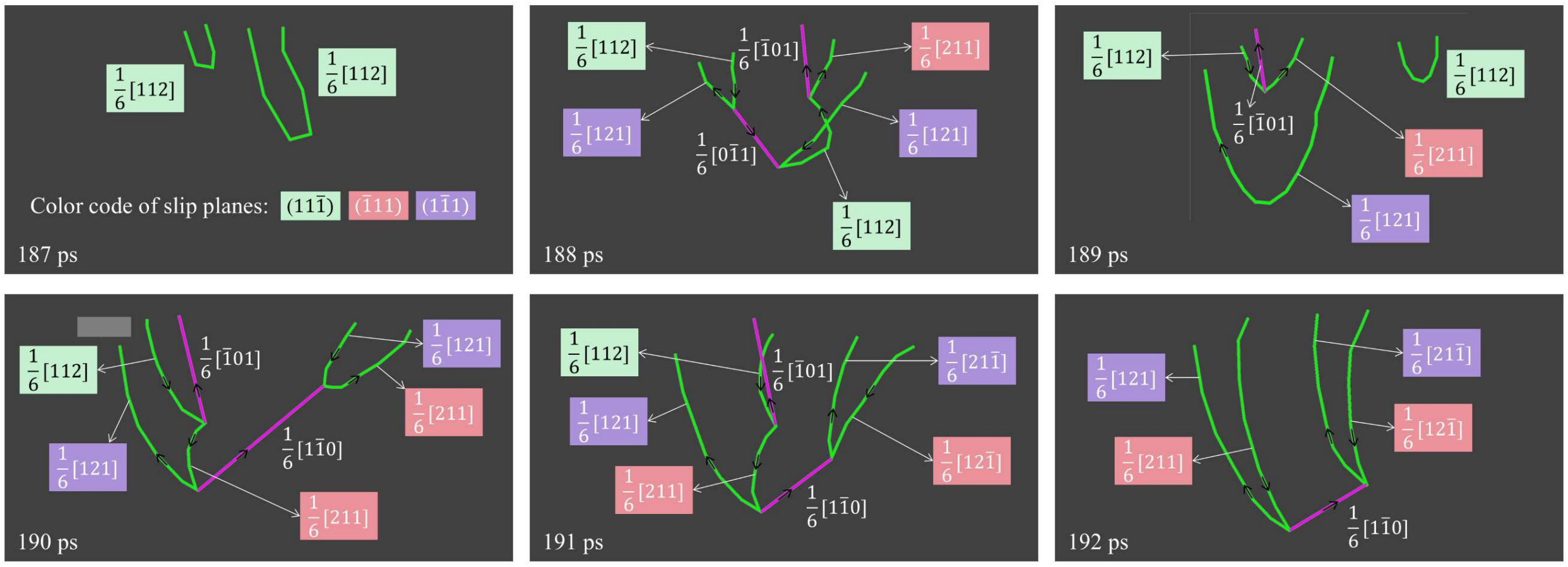

**Fig. 14.** Schematic drawing of the sequence of dislocation reactions leading to the formation of the surface-anchored "swing-like" dislocation structure marked **A** in the snapshot at $t$ = 200 ps in Fig. 13. Dislocation lines are color coded according to their types, as defined by the legend in Fig. 13. The Burgers vector of each dislocation is shown in a box with the background color indicating the corresponding slip plane, according to the color scheme defined in the panel at $t$ = 187 ps. Small black arrows along the dislocation lines indicate the line sense directions used to assign the signs of the Burgers vectors.

The cross-slip of SAW-induced Shockley partial half-loops described above is similar to the cross-slip of a leading partial in the Fleischer mechanism for cross-slip of a dissociated screw dislocation [74] and has also been observed for individual partial dislocations in MD simulations of dislocation emission from sharp cracks in an LJ material with a low stacking-fault energy [75]. In the model Pt described by the EAM potential, the normalized intrinsic stacking-fault energy is $\gamma_{ISF}/G_H b = 0.0134$, where $\gamma_{ISF}$ = 168.05 mJm$^{-2}$ [46], $G_H$ = 44.5 GPa is the Hill-averaged shear modulus, and $b = a/\sqrt{2}$ = 2.819 Å is the magnitude of the Burgers vector of a perfect dislocation. This normalized stacking-fault energy is among the higher values for fcc metals [76] and favors a relatively narrow equilibrium separation of Shockley partials, as is indeed observed in dislocation configurations that undergo gradual evolution below the surface at later times. The initial prevalence of individual Shockley partial half-loops undergoing rapid cross-slip is therefore not an artifact of an anomalously low stacking-fault energy predicted by the potential. Rather, it reflects the extreme, highly localized, and rapidly changing stresses generated at the SAW shock front.

The rapidly varying stress field behind the SAW shock front pulls some elements of the generated dislocation structures back toward the surface while driving others deeper into the substrate, as illustrated by the snapshots in Fig. 13. The formation of the largest dislocation loop, which extends to a depth of approximately 53 nm by 340 ps, is schematically illustrated in Fig. 15. At 330 ps, two dissociated $\frac{1}{2}[110]$ dislocation half-loops of opposite signs, as indicated by their opposite line senses, approach each other on the $(\bar{1}11)$ and $(1\bar{1}1)$ slip planes. By 335 ps, the two dislocations react, partially annihilate, and rearrange into two new loops, each incorporating extended segments of the original half-loops. As a result, each new loop contains segments of comparable length on the $(\bar{1}11)$ and $(1\bar{1}1)$ slip planes. The smaller loop rapidly retreats to the surface and disappears by 340 ps, whereas the larger loop remains in the substrate, with three sessile $\frac{1}{6}[1\bar{1}0]$ stair-rod segments marking the transitions between the two slip planes. As discussed in the next subsection, the presence of this dislocation structure at a depth of approximately 53 nm plays a key role in the subsequent activation of a dislocation source on the $(111)$ plane, leading to the formation of a twinned domain oriented parallel to the surface.

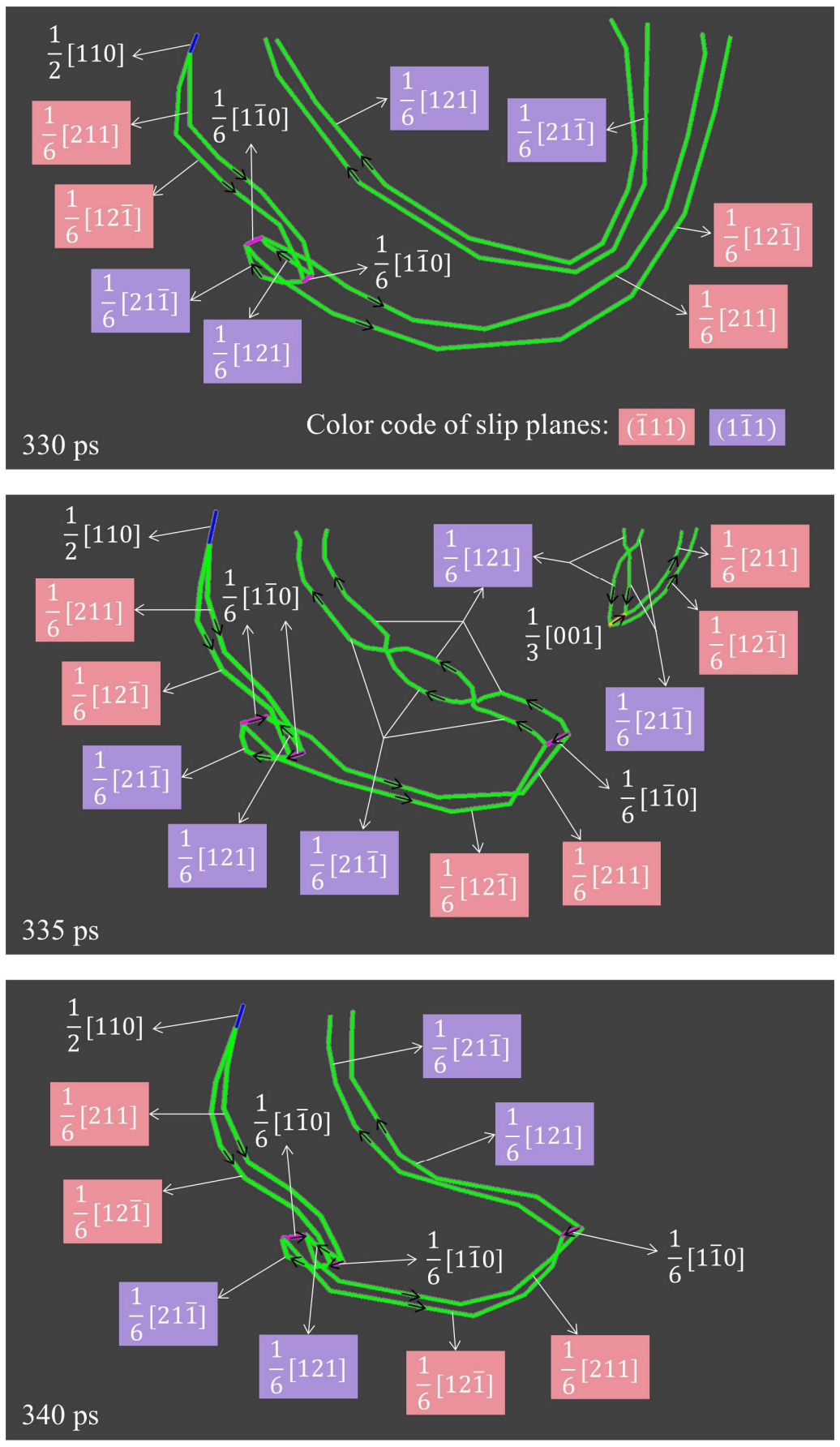

**Fig. 15.** Schematic drawing of the sequence of dislocation reactions resulting from the interaction of two $\frac{1}{2}[110]$ dislocation half-loops and leading to the formation of a dislocation half-loop composed of segments on the $(\bar{1}11)$ and $(1\bar{1}1)$ slip planes, with three transitions between the two planes, each marked by the formation of a short $\frac{1}{6}[1\bar{1}0]$ dislocation segment. By $t$ = 340 ps, the loop extends to a depth of approximately 53 nm below the surface and is marked **B** in the corresponding snapshot in Fig. 13. Dislocation lines are color coded according to their types, as defined by the legend in Fig. 13. The Burgers vector of each dislocation is shown in a box with the background color indicating the corresponding slip plane, according to the color scheme defined in the panel at $t$ = 330 ps. Small black arrows along the dislocation lines indicate the line sense directions used to assign the signs of the Burgers vectors.

### *3.2.3. SAW-driven nonthermal annealing of dislocations*

The generation of dislocations at the SAW shock front at about 180 ps coincides with a marked decrease in the peak RSS (Figs. 11b and 12), suggesting that the onset of plastic deformation opens an additional channel for dissipation of strong SAWs. To examine the subsequent interaction of the gradually weakening SAW with the defect structures generated during its initial propagation through the system, the simulation is continued as the wave reenters through the periodic boundary and repeatedly traverses the computational system. With the system length along the propagation direction $L_z$ = 244 nm and the linear SAW velocity $c$ = 1278 m/s (Table A1), the wave circulation period is approximately 191 ps. Since the first dislocations are generated at about 180 ps, the reentering shock front reaches the structurally modified region at approximately 370 ps. Owing to the depth dependence of the SAW stress field (Fig. 2), the shock front arrives somewhat earlier in deeper regions of the substrate and begins to interact with the generated dislocation structures at approximately 360 ps.

Although the shock amplitude progressively decreases due to wave dissipation, transient emission of dislocation half-loops at the shock front can still be observed in the snapshots in Fig. 16 up to approximately 600 ps, during the second and third passages of the SAW through the structurally modified material. These newly generated dislocations, however, do not expand far into the substrate and rapidly retreat to the surface as the RSS on the corresponding slip systems reverses sign behind the shock front (Fig. 12). More importantly, the repeated passages of the SAW drive substantial evolution of the subsurface dislocation structures generated during the preceding passages.

This interaction is vividly illustrated by the evolution of the large dissociated $\frac{1}{2}[110]$ dislocation half-loop formed through the dislocation reactions discussed in Section 3.2.2 and shown in Fig. 15. As the returning SAW shock front approaches the half-loop, the RSS acting on its segments on the $(\bar{1}11)$ and $(1\bar{1}1)$ slip planes increases strongly in the upper part of the loop, where nonlinear SAW sharpening concentrates elastic energy and stress near the surface. The increase in the near-surface RSS on these slip systems during shock formation is evident in Figs. 11c and 12. In contrast, at a depth of approximately 60 nm, the RSS on the same slip systems is much less affected by nonlinear sharpening and remains below 1 GPa until the returning SAW reaches this region at about 360 ps (Fig. 17a). Under the strong shear stresses generated by the returning SAW in the near-surface region, the two sides of the upper part of the $\frac{1}{2}[110]$ half-loop, located on the $(\bar{1}11)$ and $(1\bar{1}1)$ slip planes, are driven toward each other and partially annihilate at approximately 372 ps, splitting the half-loop into two parts. The upper part retreats toward the surface and eventually disappears, whereas the lower part decomposes into several dislocation structures that continue to evolve under the rapidly varying stress field of the propagating SAW.

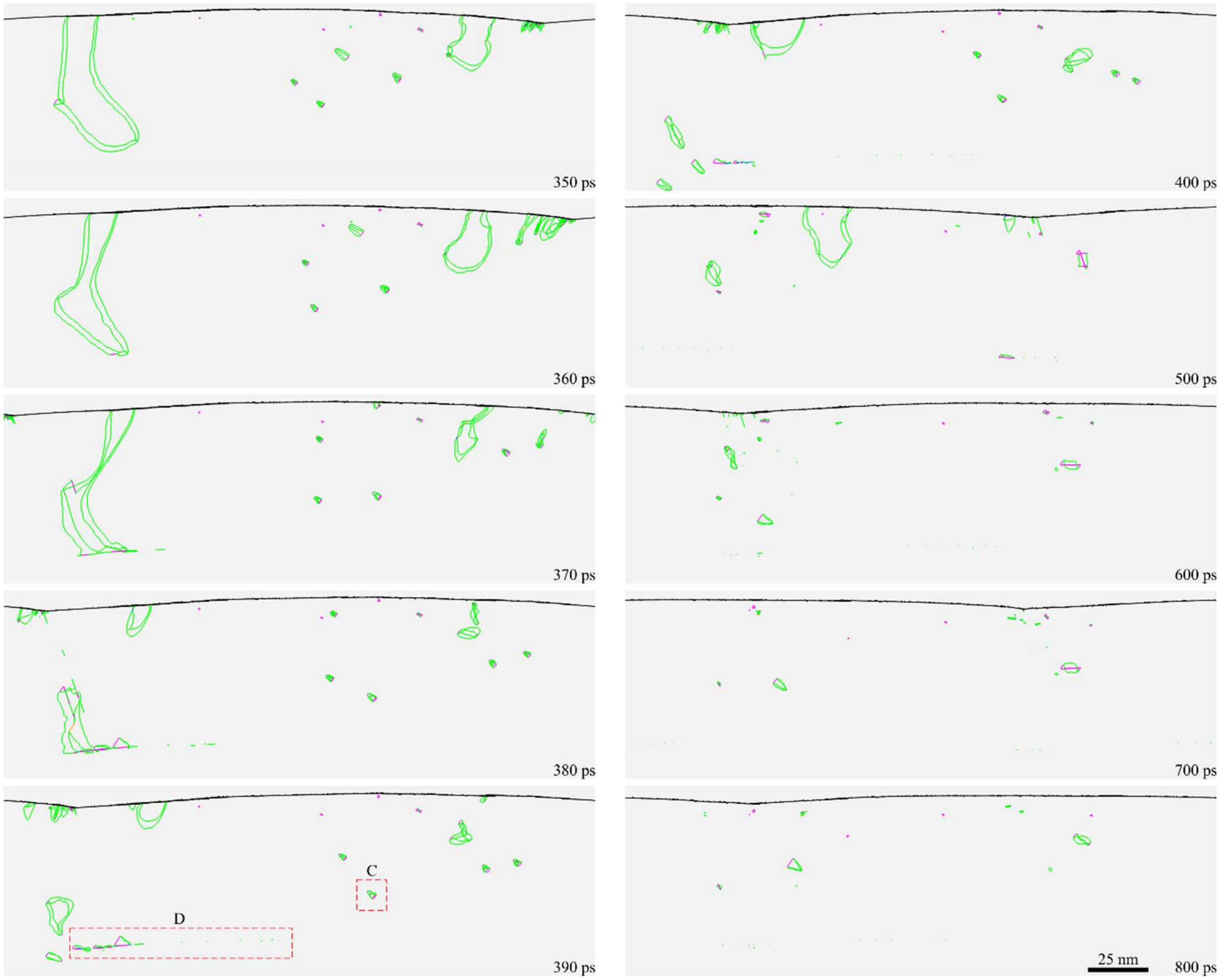

**Fig. 16.** Snapshots illustrating the evolution of the dislocation structures generated following shock front formation and illustrated in Fig. 13 under repeated passages of the SAW through the system. The snapshots demonstrate a progressive reduction in the dislocation density during successive passages of the SAW. The displayed domain spans the full system length along the propagation direction and extends to a depth $\Delta z = 78$ nm. The color coding of dislocation types is the same as in Fig. 13. The dislocation structure outlined by box **C** and the formation of a twinned domain in the region outlined by box **D** in the panel at $t = 390$ ps are detailed in Fig. S3 of the Supplementary Material and Fig. 17, respectively.

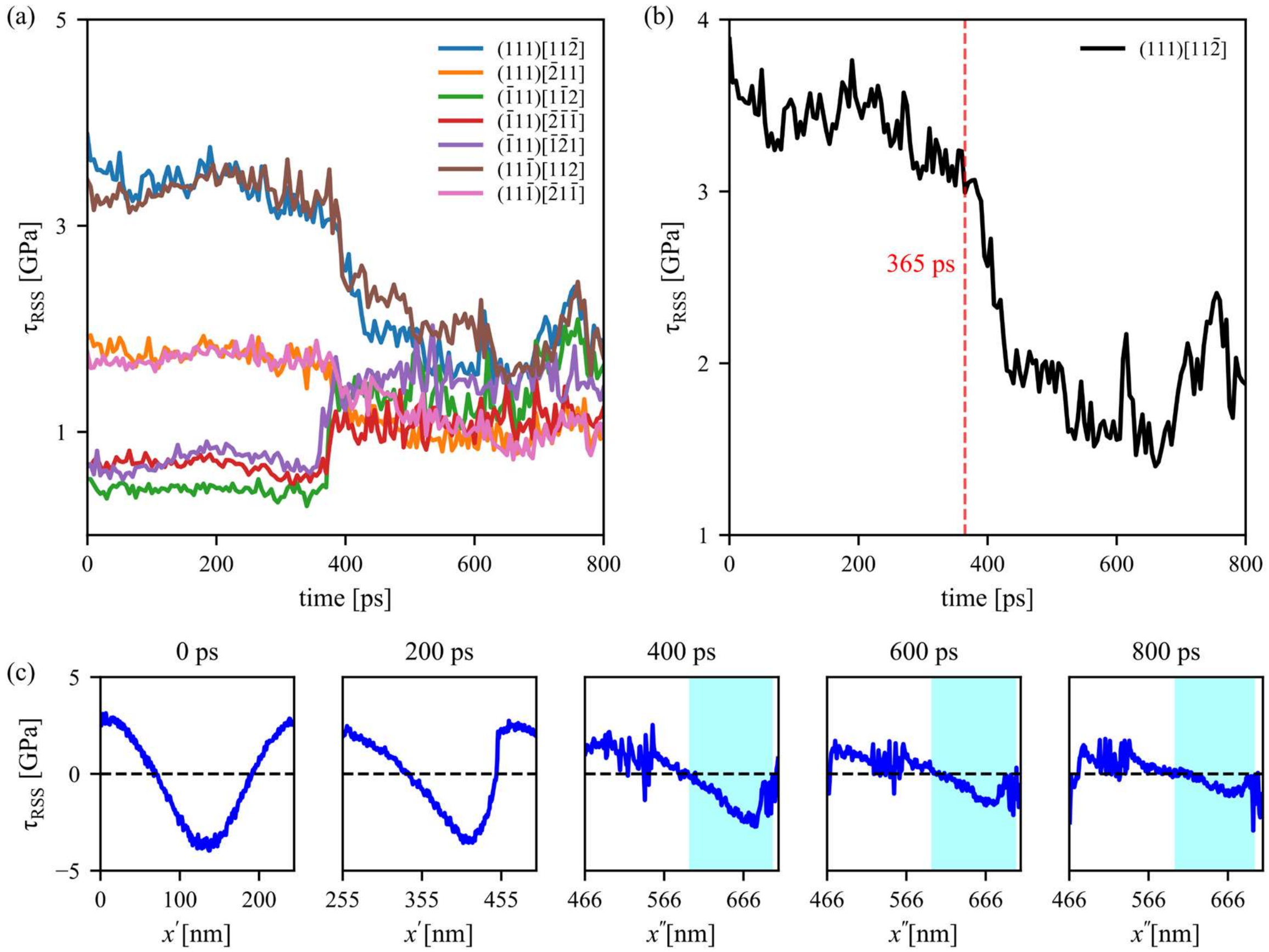

**Fig. 17.** Conditions leading to the formation of the twinned domain at a depth of approximately 60 nm below the surface illustrated in Fig. 18. (a) Comparison of the time evolution of the peak RSS values for all Shockley partial slip systems at a depth of 60 nm ($z$ = -60 nm), with only one representative system shown for each symmetry-equivalent set. (b) Evolution of the peak RSS for the $(111)[11\bar{2}]$ slip system on which a source of Shockley partial dislocations leading to the formation of the twinned domain is activated. The time of 365 ps, when the first Shockley partial is emitted from the source, is marked by the vertical red dashed line. (c) Evolution of the RSS profile for the $(111)[11\bar{2}]$ slip system during SAW propagation. The region in which the twinned domain forms is highlighted in light blue background. For the 0 and 200 ps panels, the viewing window follows the SAW, with the horizontal coordinate defined as $x' = x + ct$, where $c$ is the linear SAW velocity. Once the SAW shock front reaches the region containing dislocations generated during the preceding cycle, the viewing window is fixed relative to the substrate, with the horizontal coordinate defined as $x'' = x + ct_0$, where $t_0$ = 365 ps is the time when the first Shockley partial dislocation is emitted from a source located at a depth of approximately 60 nm on the $(111)[11\bar{2}]$ slip system. The panels at 400, 600, and 800 ps therefore illustrate successive passages of the SAW through the same region of the substrate.

While some of these structures rearrange and gradually glide toward the surface, one configuration acts as a source for the emission of additional dislocations on the $(111)[11\bar{2}]$ slip system. As shown in Fig. 17a, the RSS on this system remains at approximately 3.5 GPa at a depth of 60 nm during wave sharpening, compared with about 0.5 GPa near the surface (Fig. 11c). Together with the RSS on the $(11\bar{1})[112]$ system, these are the highest values observed among the Shockley partial slip systems at this depth. Importantly, the peak RSS on the $(11\bar{1})[112]$ system also remains largely unchanged at this depth during wave sharpening. This behavior differs from that near the surface, where the peak RSS on the same slip system increases substantially as the shock front develops and ultimately reaches the level required for dislocation emission (Figs. 11c and 12). The difference reflects the progressive localization of elastic energy and associated stresses near the surface during nonlinear SAW sharpening. At a depth of 60 nm, the RSS profiles still undergo pronounced sharpening, as illustrated for the $(111)[11\bar{2}]$ system in Fig. 17c, but this sharpening does not produce a corresponding increase in the peak stress.

The peak RSS of approximately 3.5 GPa is apparently insufficient for homogeneous nucleation of dislocation loops, and no dislocations are generated at this depth during the initial formation of the SAW shock front. Once dislocations generated near the surface extend to this depth, however, they provide heterogeneous sources that substantially reduce the barrier for further dislocation emission. The source that develops at a depth of approximately 60 nm consists of a complex arrangement of Shockley partial, stair-rod, and Frank dislocations and sequentially emits multiple Shockley partials with the same $\frac{1}{6}[11\bar{2}]$ Burgers vector. The newly emitted dislocations have predominantly edge character, are largely aligned along the $[1\bar{1}0]$ direction, and glide along $[11\bar{2}]$, parallel to the SAW propagation direction. Their glide on successive $(111)$ planes produces an expanding twinned domain oriented parallel to the surface. The deformation-twinning process [77] is illustrated in Fig. 18a, where atoms within the expanding twin lamella are highlighted in yellow and the cores of the Shockley partials responsible for twin growth are marked by the symbols $\perp$ and $\top$.

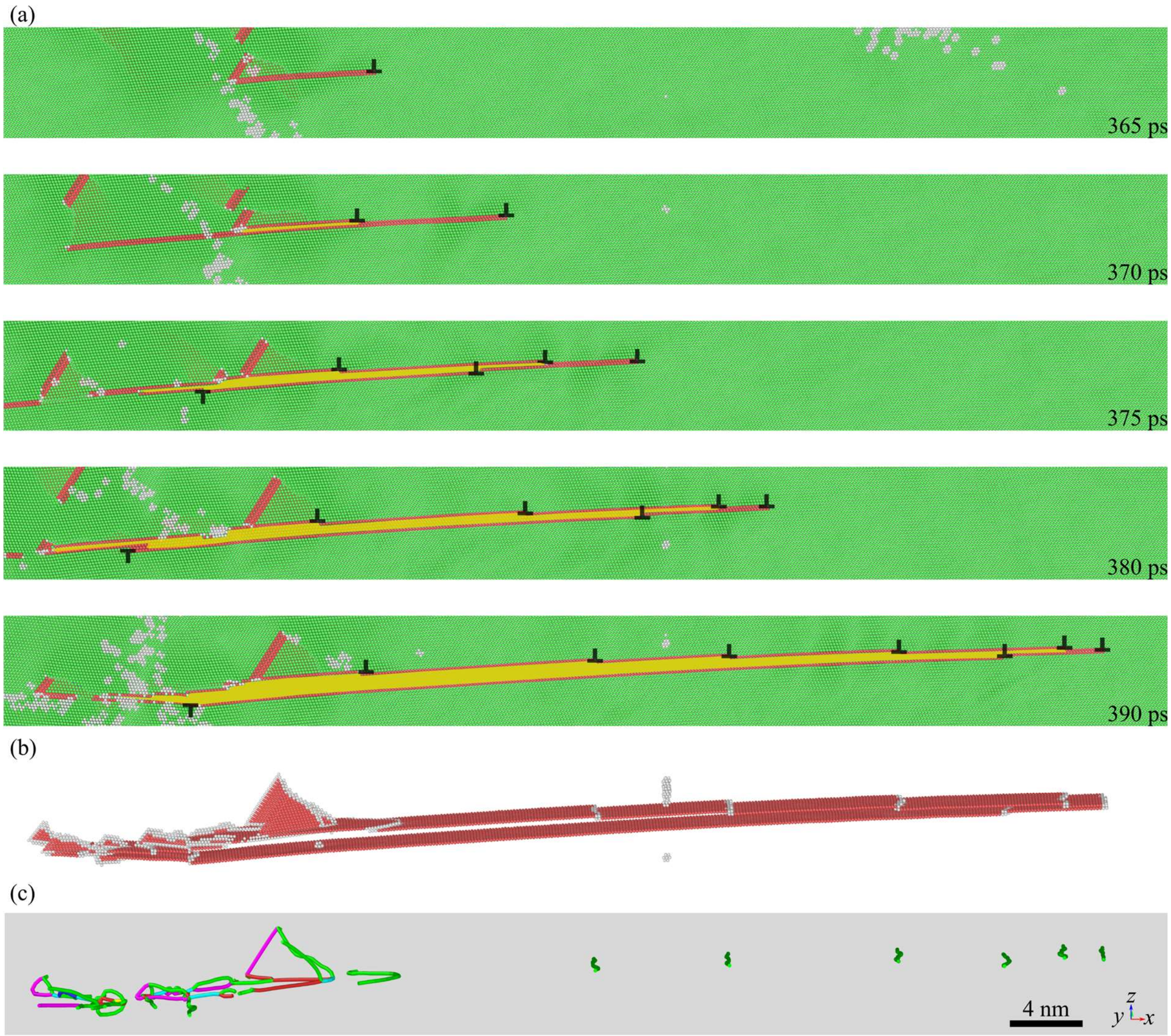

**Fig. 18.** SAW-driven formation of a twinned domain oriented parallel to the surface at a depth of approximately 60 nm. (a) Cross-sections of atomic configurations at selected times from $t$ = 365 to 390 ps. Atoms within the expanding twinned domain are colored yellow, twin boundaries and stacking faults are colored red, and the cores of Shockley partial dislocations with edge character are denoted by the symbols ⊥ and ⊤. (b) Atomic configuration at $t$ = 390 ps showing only atoms associated with twin boundaries, stacking faults, and dislocation cores. (c) Dislocation structure at $t$ = 390 ps, with the dislocation types color coded according to the legend in Fig. 13. The views in (b) and (c) are slightly rotated about the $x$-axis to better reveal the dislocations and stacking faults. The location of the twinned domain is outlined by box **D** in Fig. 16.

Similar to the RSS responsible for the emission and subsequent evolution of dislocation half-loops in the near-surface region, the RSS acting on the $(111)[11\bar{2}]$ slip system at a depth of 60 nm changes sign as the SAW propagates through the region containing the evolving twinned domain. The magnitude of the RSS decreases markedly during the initial growth of the twin, falling below 2 GPa by 420 ps (Fig. 17b). Subsequent passages of the weakened SAW through this region lead to a reduction in the length of the twinned domain by approximately 28% from its maximum value by 600 ps and by an additional 7% by 800 ps. This evolution is illustrated in Fig. 17c, where the extent of the twinned domain is indicated by the light blue shading.

The activation of additional slip systems and formation of the twinned domain demonstrate that repeated SAW interactions with an existing dislocation structure can locally generate new defects. Nevertheless, the overall evolution of the dislocation configurations in Fig. 16 reveals a different net trend. Once the returning SAW begins to propagate through the defect-containing region at approximately 360 – 370 ps, dislocations undergo repeated rearrangement, glide, reactions, and annihilation, with many segments ultimately retreating to the free surface. The resulting progressive reduction in the overall dislocation density is already apparent from the snapshots in Fig. 16. By 800 ps, after more than three passages of the SAW through the structurally modified region, the number and extent of the remaining dislocation structures are substantially reduced. Thus, interaction of the SAW with a defect-containing crystal promotes defect relaxation and removal, in contrast to the net defect generation that occurs when the newly formed SAW shock front propagates through the initially defect-free crystal.

This change in the role of the SAW is quantified in Fig. 19. The dislocation density averaged over the top 50 nm of the substrate is calculated at intervals of 100 ps and plotted in Fig. 19a. Following the onset of dislocation nucleation at the shock front at approximately 180 ps, the dislocation density increases rapidly as the shock front propagates through the initially defect-free region. The depth profile at 300 ps (Fig. 19b) shows that the generated dislocations are largely confined within approximately 30 nm of the surface, with Shockley partials accounting for the majority of the total dislocation length. The initial increase in dislocation density is followed by a steady decrease once the SAW begins to interact with the defect-containing region at approximately 360 ps. The simulations therefore reveal two distinct regimes of structural modification by strong SAWs: defect generation during shock propagation through an initially defect-free crystal and nonthermal annealing during subsequent interactions of a weaker SAW with the generated defect structures. The SAW-driven rearrangement and removal of dislocations observed here may be related to the broader phenomenon of acoustic modification of dislocation structures observed under ultrasonic excitation [78,79].

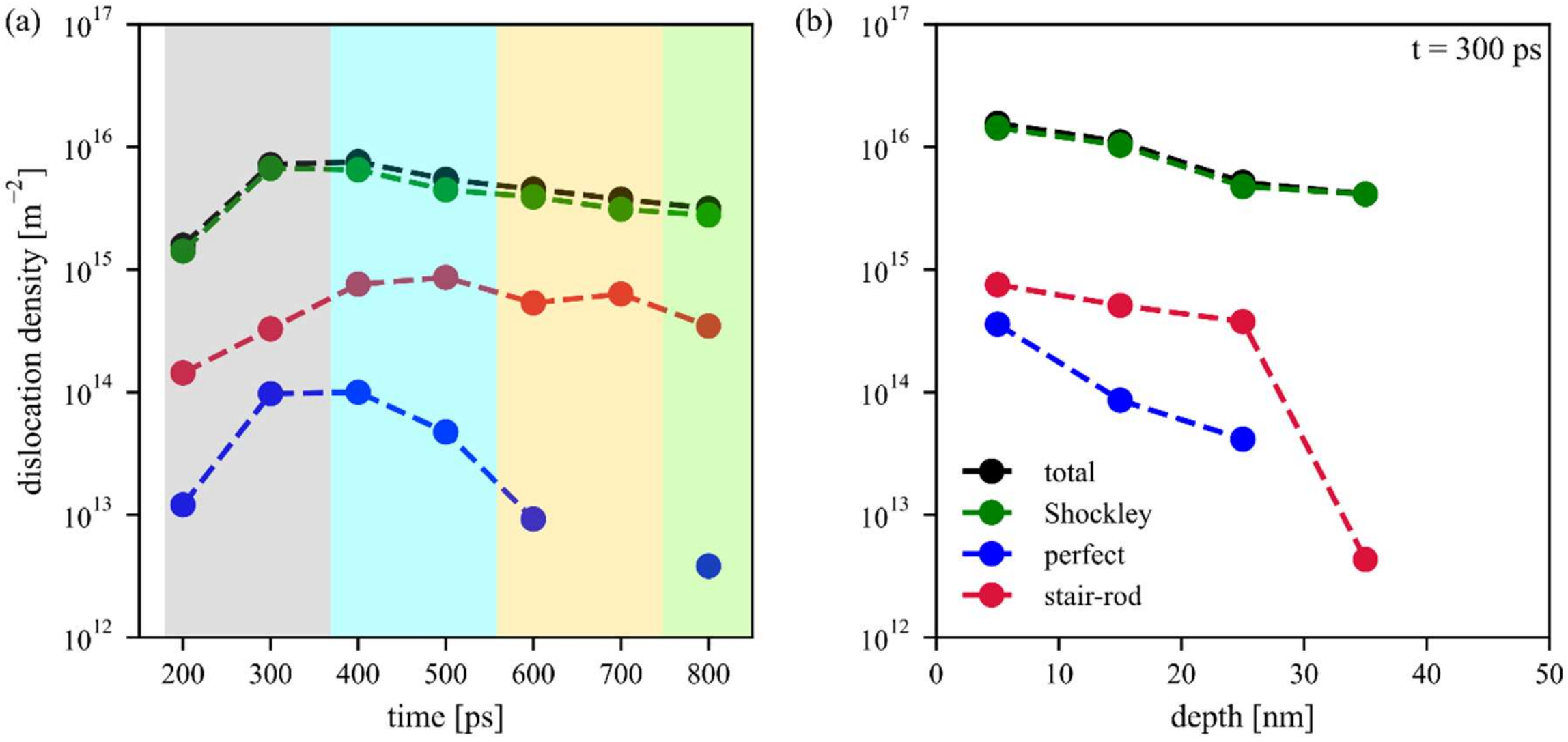

**Fig. 19.** Spatial distribution and temporal evolution of the dislocation density induced by a SAW with an initial strain amplitude $\varepsilon_{xx}^0$ of 4% on the Pt(111) surface. (a) Temporal evolution of the dislocation density in the near-surface region, obtained by averaging over the top 50 nm of the substrate. Different background shades indicate successive passages of the SAW through the system, with the passage intervals counted from the time of the first dislocation emission at the shock front. (b) Dislocation density as a function of depth below the surface at $t$ = 300 ps. The data points are obtained by averaging the dislocation density over 10-nm-thick layers. Zero values are omitted from the plots, with connecting lines interrupted at the corresponding positions.

### *3.3. Atomic-scale surface roughness*

An important implication of SAW-induced dislocation generation is the atomic-scale modification of the Pt surface structure. Dislocation emission from the free surface produces atomic steps with heights determined by the Burgers vectors of the emitted dislocations. The connection between dislocation emission and the development of atomic-scale surface morphology is illustrated in Fig. 20, which shows the atomic structure of a small region of the top surface monolayer as the SAW shock front passes through this region. The surface is viewed from the substrate interior, with atoms colored according to their potential energy and dislocation half-loops originating at the surface shown as green and pink lines for Shockley and stair-rod dislocations, respectively. The shock front propagates from left to right, and atoms behind the shock front have lower average potential energies, consistent with the sharp decrease in the normal stress $\sigma_{xx}$ and strain $\varepsilon_{xx}$ across the shock front (Figs. 3 and 6). More importantly, the atomic steps connecting the points where the dislocations intersect the surface are marked by rows of atoms with distinctly different potential energies. The red atoms displaced upward by the dislocations have higher potential energies, reflecting their less favorable local atomic environment, whereas the blue atoms displaced downward have lower potential energies.

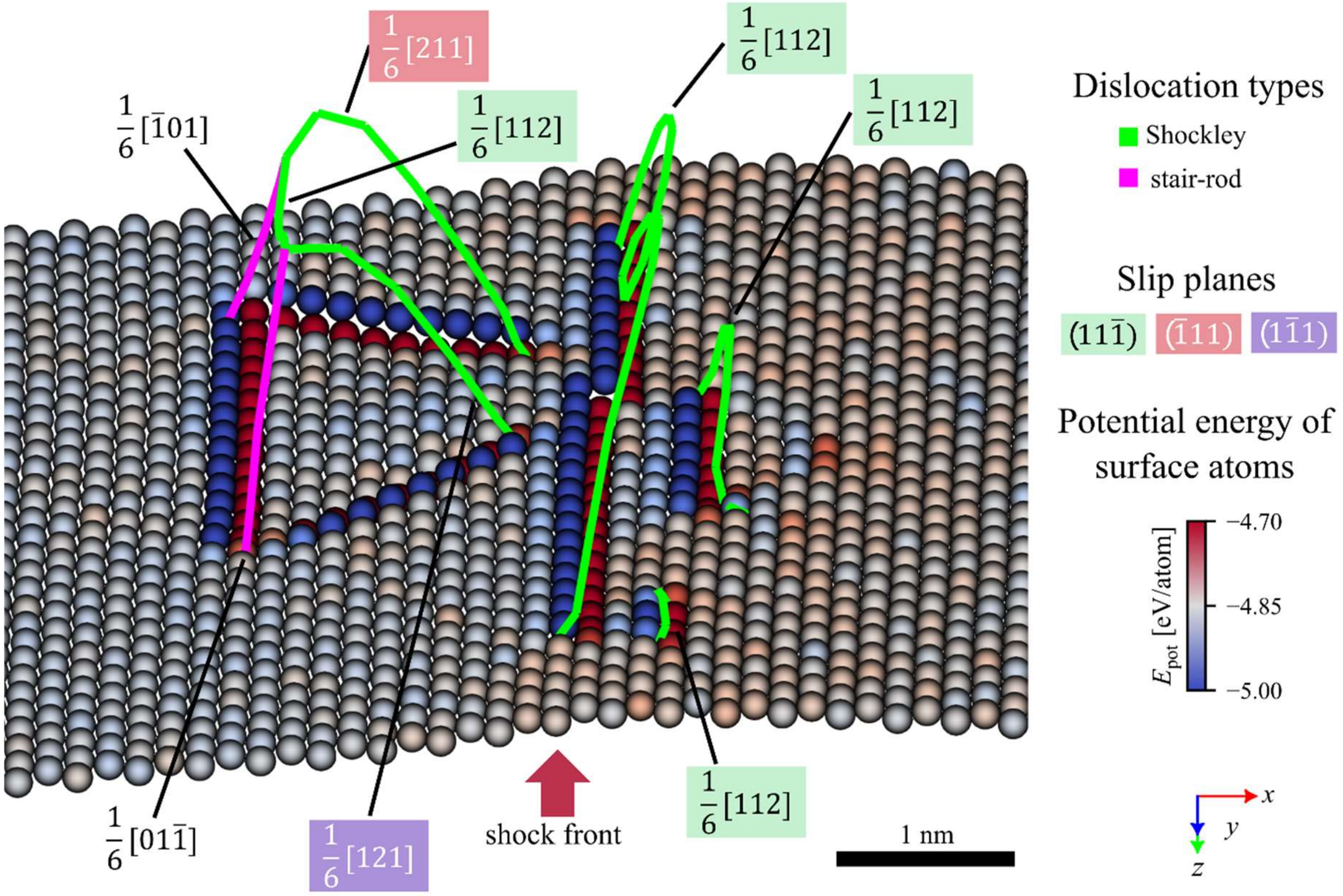

**Fig. 20.** Example of a near-surface dislocation configuration observed at *t* = 185 ps in the vicinity of the SAW shock front, illustrating the connection between dislocation emission and the generation of atomic-scale surface roughness. To reduce the thermal noise in atomic positions and energies, the atomic configuration is quenched for 0.5 ps using a velocity-dampening technique. Only atoms in the top monolayer of the substrate are shown and are colored according to their potential energy. Atoms behind the shock front (in the left part of the snapshot) have higher potential energies. The surface is viewed from the substrate interior toward the surface.

The evolution of the atomic-scale surface morphology is illustrated in Fig. 21, which shows top views of a representative surface region colored according to atomic potential energy at selected times during SAW propagation. No surface defects are visible in this region at 200 ps, even though the entire SAW waveform has already passed through it. By 300 ps, however, multiple rows of high-energy atoms, individual surface vacancies and adatoms, and small clusters of these point defects have appeared. The linear rows of high-energy atoms can be directly associated with atomic steps produced by surface-terminated dislocation loops, as also illustrated in Fig. 20. The formation of vacancies, vacancy clusters, and adatoms is likely facilitated by the transient emergence and disappearance of dislocation loops at the shock front, where the formation and removal of atomic steps, combined with the large local strains, create favorable conditions for the generation of surface point defects and their clusters.

As the evolution proceeds to 400 and 600 ps, some dislocations retract to the surface, eliminating the corresponding atomic steps, while new dislocations continue to be generated at the shock front and subsequently retract to the surface behind it. The orientations of the atomic steps reflect the activation of three slip planes. Dislocations emitted on the $(11\bar{1})$ planes produce steps oriented perpendicular to the SAW propagation direction, whereas dislocations emitted on the $(1\bar{1}1)$ and $(\bar{1}11)$ planes produce the two sets of inclined steps.

By 800 ps, the weakened SAW no longer generates new dislocations at the shock front, and all surface-terminated dislocation loops have retracted to the surface, as seen in Fig. 16. Consequently, the extended high-energy atomic rows associated with these dislocations have largely disappeared. The surface nevertheless retains numerous atomic-scale features, including small clusters of surface point defects, larger terraces and islands, and short rows of high-energy atoms. Thus, the transient dislocation activity leaves a persistent population of atomic-scale surface defects even after the surface-terminated dislocations have disappeared.

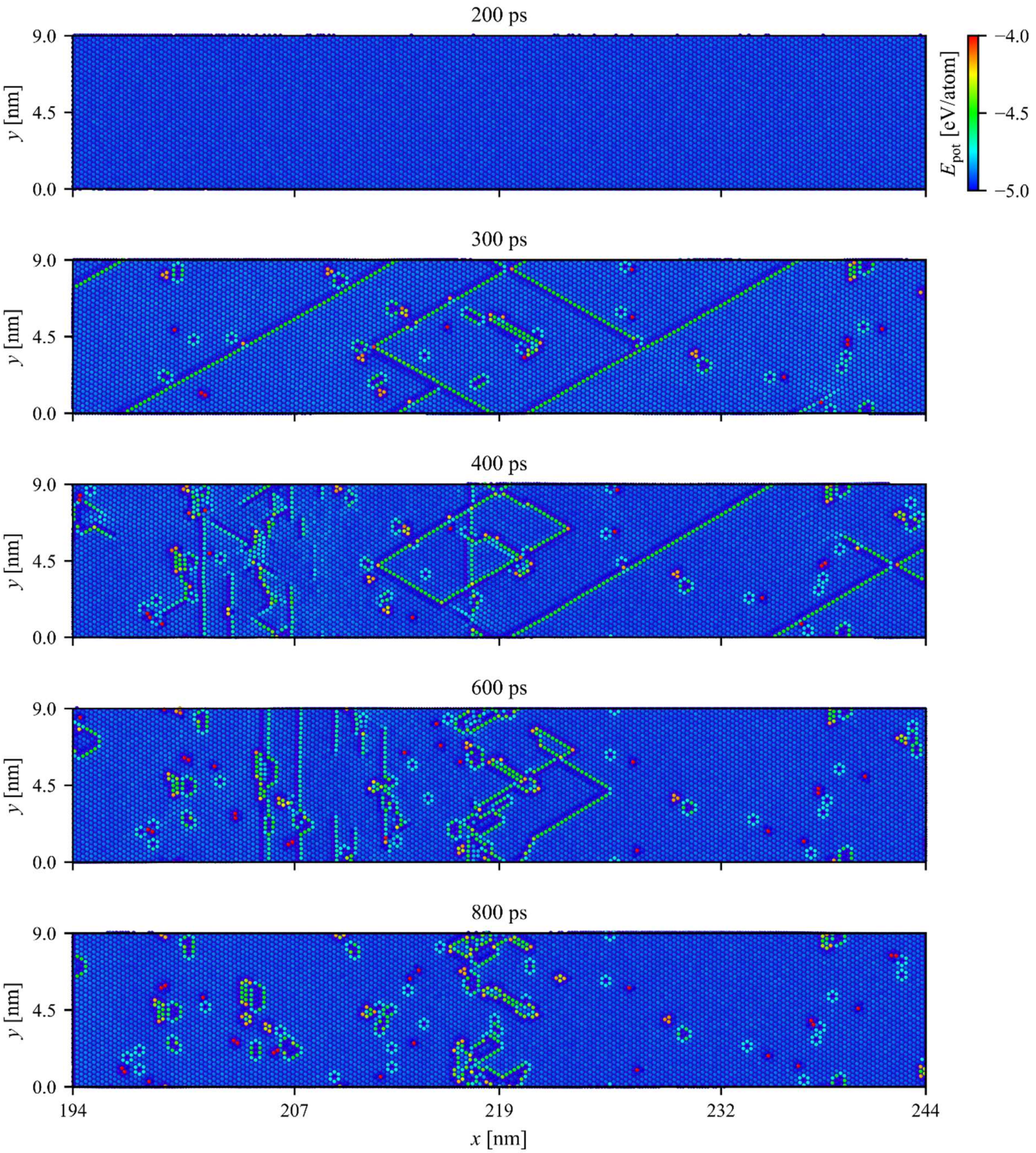

**Fig. 21.** Top view of a representative surface region colored according to atomic potential energy at selected times during SAW propagation on the Pt(111) surface with an initial strain amplitude of $\varepsilon_{xx}^{0} = 4\%$. To reduce the thermal noise in atomic positions and energies, the atomic configurations are quenched for 0.5 ps using a velocity-dampening technique. The generation and evolution of dislocations in the near-surface region produce atomic-scale surface features such as steps, terraces, and islands.

To provide a quantitative description of the SAW-induced evolution of the atomic-scale surface morphology, the coordination numbers (CNs) of atoms in the topmost surface layer are analyzed in Fig. 22. The CN of each atom is determined by counting its neighbors within a cutoff distance of 1.1 times the nearest-neighbor distance in Pt at 300 K. Atoms on an ideal (111) surface have a CN of 9, and the appearance of undercoordinated surface atoms with CN < 9 provides a measure of atomic-scale surface roughening. During the initial propagation of the SAW through the pristine crystal, the first undercoordinated atoms appear at 180 ps, coinciding with the first dislocation emission (Section 3.2.1). Their fraction is still low (0.92 %) at 200 ps, when only a relatively small part of the system has been affected by dislocation activity (Fig. 13). The fraction of undercoordinated atoms increases sharply by 300 ps and reaches its maximum by 400 ps. At later times, it decreases slightly and stabilizes at approximately 7% (7.64% at 600 ps, 7.22% at 700 ps, and 7.08% at 800 ps).

The evolution of the surface structure broadly follows that of the subsurface dislocation structures, with the sharp increase in the fraction of undercoordinated atoms occurring during active emission of surface-terminated dislocation loops at the SAW shock front and the subsequent decrease coinciding with SAW-driven nonthermal annealing of the dislocation structures. Although the dislocation density in Fig. 19a and the fraction of undercoordinated surface atoms in Fig. 22 exhibit similar initial increases and both reach their maxima at approximately 400 ps, their subsequent declines differ substantially. Between 400 and 800 ps, the dislocation density decreases by a factor of 2.4, whereas the fraction of undercoordinated surface atoms decreases by only 18%, from 8.63% to 7.08%. Thus, the nonthermal annealing of the subsurface dislocation structures does not lead to a corresponding recovery of the surface. Instead, much of the atomic-scale surface modification generated during the period of intense dislocation activity persists even after the subsurface dislocation density has substantially decreased.

The distribution of undercoordinated atoms among different coordination numbers, shown by the stacked bars in Fig. 22a, indicates that atoms with CN = 7 and 8 dominate the undercoordinated population. Most of these atoms are associated with atomic steps generated by the emission of dislocation half-loops, as well as with individual surface vacancies and small vacancy clusters. The subsequent retraction of dislocation loops to the surface reduces the relative abundance of atoms with CN = 7 and 8, which stabilizes at approximately 80% of the undercoordinated population by 800 ps. In contrast, the relative fraction of atoms with lower coordination numbers (CN = 3–6) increases during the first SAW circulation following shock formation and the onset of dislocation emission, and remains nearly constant after 400 ps. These atoms occupy a variety of surface configurations, including isolated adatoms (CN = 3), adatom pairs (CN = 4), adatom triplets (CN = 5), and kinks and corners of atomic islands and depressed terraces (CN = 6).

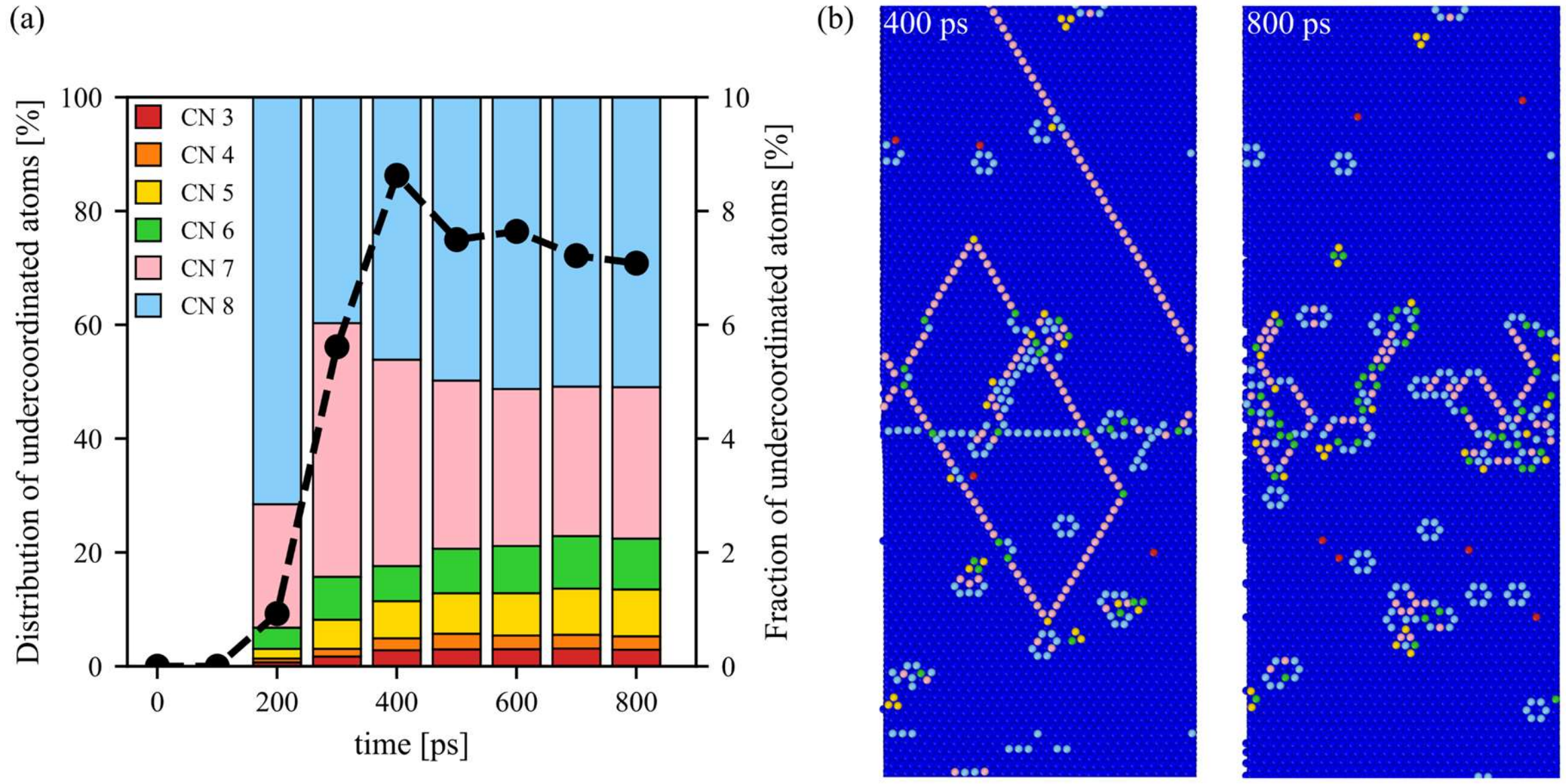

**Fig. 22.** Coordination analysis of the topmost surface monolayer during SAW propagation on the Pt(111) surface with an initial strain amplitude of $\varepsilon_{xx}^{0} = 4\%$. (a) Temporal evolution of the fraction of undercoordinated atoms on the Pt(111) surface and their distribution among different coordination numbers (CNs). The black dashed line with data points shows the fraction of atoms with CN = 3–8 (undercoordinated atoms) relative to the total number of atoms with CN = 3–9, with CN = 9 corresponding to atoms on an ideal Pt(111) surface. The stacked bars show the relative fractions of undercoordinated atoms with CN = 3–8, normalized by the total number of undercoordinated atoms at each time. (b) Top view of two representative surface regions at 400 and 800 ps, with atoms colored according to their coordination number using the same color scheme as the stacked bars in panel (a). Atoms with CN = 9, corresponding to the coordination number of atoms on an ideal (111) surface, are shown in dark blue. A larger surface region at additional times is shown in Fig. S4 of the Supplementary Material.

Experimental evidence for atomic-scale surface restructuring without melting has also been reported in studies of metal surfaces subjected to low-fluence femtosecond laser irradiation. Scanning tunneling microscopy studies revealed the formation of adatom and vacancy clusters following exposure to repeated ultrashort laser pulses at fluences far below the melting and ablation thresholds [ 80 - 82 ]. These modifications were attributed to nonthermal processes involving the generation of subsurface point defects and their transport to the surface, followed by surface diffusion and clustering. Although the underlying mechanisms differ from the mechanically driven dislocation activity observed in the present simulations, both types of excitation demonstrate the possibility of substantial atomic-scale surface restructuring without melting.

The ability of strong SAWs to produce or substantially increase atomic-scale surface roughness may be relevant to the persistent observations of enhanced catalytic activity and selectivity of surfaces subjected to SAWs [34-39], for which a conclusive mechanistic explanation is still lacking. A recent critical analysis of possible promotion mechanisms [40] concluded that electronic effects associated with the transient modification of the local work function by SAW-induced strain are too small to make a significant contribution. Among the mechanisms considered more plausible are temperature variations associated with SAW propagation and dissipation, dynamic coupling to nonlinear components of the SAW [23], and reversible changes in the atomic structure of the surface. The possible role of structural modification is also supported by observations of mechanical damage to Pt catalyst films subjected to SAWs [37].

The present results provide a microscopic basis for the proposed contribution of SAW-induced structural modification to enhanced catalytic activity. Atomic steps, vacancies, adatoms, and other atomic-scale defects on Pt surfaces are well known to modify catalytic activity [83,84], with particular atomic configurations providing active sites for specific catalytic reactions [84-86]. The simulations demonstrate that nonlinear SAW propagation can generate a high density of atomic-scale surface defects through dislocation emission and subsequent evolution of the dislocation structures. Importantly, many of these surface features persist even as the subsurface dislocation density decreases through SAW-driven nonthermal annealing. The resulting dynamic generation, restructuring, and partial retention of atomic-scale surface defects provide a plausible microscopic pathway through which strong SAWs may modify the catalytic activity of metal surfaces.

## 4. Summary

Large-scale MD simulations have been used to investigate the nonlinear evolution of Rayleigh surface acoustic waves and the resulting generation and evolution of crystal defects and atomic-scale surface roughness in Pt. Simulations were performed for SAWs propagating along the $[100]$ direction on the $(001)$ surface and along the $[11\bar{2}]$ direction on the $(111)$ surface. The results reveal a strong dependence of the nonlinear SAW evolution on crystallographic orientation. On the $\mathrm{Pt}(111)$ surface, nonlinear frequency up-conversion produces rapid wave sharpening and the formation of a shock front, accompanied by progressive localization of the elastic energy and increasing stress amplitudes in a thin near-surface region. In contrast, nonlinear sharpening is much weaker for propagation on the $\mathrm{Pt}(001)$ surface, while the observed wave dissipation is expected to substantially reduce the SAW amplitude before a well-defined shock front can develop.

The formation of a shock front on the $\mathrm{Pt}(111)$ surface creates the conditions for surface nucleation of dislocations. For sufficiently strong SAWs, Shockley partial dislocation half-loops are emitted from the surface as the resolved shear stress at the shock front approaches the theoretical shear strength of a perfect crystal. The maximum RSS associated with the onset of dislocation emission is 4.99 GPa, corresponding to 0.17 and 0.13 of the relaxed and unrelaxed $\{111\}$ shear moduli, respectively. Continued propagation of the shock front activates multiple slip systems and produces complex dislocation structures through interactions and reactions among Shockley partials. As a distinctive consequence of the highly localized and rapidly changing stress field at the SAW shock front, frequent changes of the active slip planes lead to cross-slip of individual Shockley partials through a mechanism similar to the Fleischer mechanism, while interactions between partials gliding on intersecting slip planes provide an additional pathway for dislocation reactions. Both processes generate sessile stair-rod dislocation segments that temporarily pin the evolving dislocation structures.

The subsequent interaction of the SAW with the generated defects leads to further restructuring of the subsurface dislocation network. Dislocations extending from the surface into deeper regions of the substrate provide heterogeneous sources for activation of slip systems that are not activated by the propagating SAW in the initially defect-free crystal. In particular, the interaction of the SAW with dislocations extending to a depth of approximately 60 nm activates the emission of Shockley partials on $(111)$ planes parallel to the surface and leads to the formation of a subsurface twinned domain. At longer times, repeated passages of the SAW through the modified region progressively reduce the dislocation density, demonstrating a mechanism of SAW-driven nonthermal annealing of crystal defects.

The dislocation activity also produces permanent modification of the initially flat Pt surface. The emergence and disappearance of dislocations at the free surface generate atomic-scale steps, terraces, islands, and other under-coordinated atomic configurations, resulting in progressive development of atomic-scale surface roughness. These structural changes provide a possible microscopic connection between nonlinear SAW propagation and experimentally observed acoustic enhancement of surface processes, including catalytic reactions on Pt. More broadly, the ability of nonlinear SAWs to both generate crystal defects under the extreme localized stresses at the shock front and subsequently reorganize or eliminate them through continued interaction with the propagating wave demonstrates a pathway for nonthermal acoustic processing of near-surface microstructure. This perspective extends the role of SAWs beyond their more conventional use as probes of material properties and microstructure and points to their potential use for controlled modification of surface and subsurface structure.

## Declaration of Competing Interest

The authors declare that they have no known competing financial interests or personal relationships that could have appeared to influence the work reported in this paper.

## Acknowledgments

Financial support for this work was provided by the National Science Foundation (NSF) through Grant OISE-2406599. Computational support was provided by the NSF through the Advanced Cyberinfrastructure Coordination Ecosystem: Services & Support (ACCESS) project DMR110090 and the University of Virginia Advanced Research Computing Services.

## Appendix A. Analytical solution for harmonic Rayleigh wave propagation in the $[11\bar{2}]$ direction on the $(111)$ surface of a cubic crystal

The analytical solution for the displacement and velocity fields of a Rayleigh wave propagating along the $[11\bar{2}]$ direction on the $(111)$ surface of a cubic crystal is derived following the general approach used in earlier studies of SAW propagation on (001) surfaces [33,49,87]. For an anisotropic elastic medium, the equation of motion can be written in Einstein summation notation as

$$\rho \frac{\partial^2 u_i}{\partial t^2} = C_{ijlm} \frac{\partial^2 u_m}{\partial x_j \partial x_l}, \quad \text{(A1)}$$

where $\rho$ is density, $u_i$ is the displacement along the coordinate axis $x_i$, and $C_{ijlm}$ is the stiffness tensor. The solution of the wave equation for SAWs can be written as

$$u_j = A_j \exp[ik \cdot (x_1 cos\alpha_1 + x_2 cos\alpha_2 + x_3 cos\alpha_3 - ct)], \quad \text{(A2)}$$

where $cos\alpha_1$ and $cos\alpha_2$ define the direction of wave propagation within the surface, $cos\alpha_3$ describes the variation of the wave field normal to the surface, $k$ is the wavenumber, $c$ is the wave velocity, and $A_j$ is the displacement amplitude along the $x_j$ axis.

Substitution of the solution given by Eq. A2 to the wave equation Eq. A1 yields

$$k^2 A_m \left(C_{ijlm} cos\alpha_j cos\alpha_l - \delta_{im} \rho c^2\right) = 0. \quad \text{(A3)}$$

For a wave that propagates along $x_1$ direction on the surface, $cos\alpha_1 = 1$ and $cos\alpha_2 = 0$.

For a SAW propagating along the $[11\bar{2}]$ direction on the $(111)$ surface, the $x_3$ axis is oriented normal to the surface, while the $x_1$ and $x_2$ axes are parallel to the $[11\bar{2}]$ and $[1\bar{1}0]$ directions, respectively. The stiffness tensor is transformed to this coordinate system, yielding the following nonzero components:

$$\begin{aligned}
&C'_{11} = C'_{22} = (C_{11} + C_{12} + 2C_{44})/2,\\
&C'_{33} = (C_{11} + 2C_{12} + 4C_{44})/3,\\
&C'_{12} = (C_{11} + 5C_{12} - 2C_{44})/6,\\
&C'_{13} = C'_{23} = (C_{11} + 2C_{12} - 2C_{44})/3, \qquad \text{(A4)}\\
&C'_{44} = C'_{55} = (C_{11} - C_{12} + C_{44})/3,\\
&C'_{66} = (C_{11} - C_{12} + 4C_{44})/6,\\
&C'_{15} = -C'_{25} = -C'_{46} = -\sqrt{2}(C_{11} - C_{12} - 2C_{44})/6.
\end{aligned}$$

Using these nonzero components and denoting $cos\alpha_3 = q$ and $\rho c^2 = R$ for convenience makes it possible to rewrite Eq. (A3) for the case where wave propagates on the $(111)$ surface along the direction $[11\bar{2}]$ as

$$k^2 A_m \begin{pmatrix} C'_{11} + 2C'_{15}q + q^2 C'_{44} - R & 0 & C'_{15} + q(C'_{13} + C'_{44}) \\ 0 & C'_{66} - 2C'_{15}q + q^2 C'_{44} - R & 0 \\ C'_{15} + q(C'_{13} + C'_{44}) & 0 & C'_{44} + q^2 C'_{33} - R \end{pmatrix} = 0 \qquad \text{(A5)}$$

Since this is an equation for waves with unknown amplitudes, a nontrivial solution requires the determinant of the matrix to vanish,

$$\det \begin{vmatrix} C'_{11} + 2C'_{15}q + q^2 C'_{44} - R & 0 & C'_{15} + q(C'_{13} + C'_{44}) \\ 0 & C'_{66} - 2C'_{15}q + q^2 C'_{44} - R & 0 \\ C'_{15} + q(C'_{13} + C'_{44}) & 0 & C'_{44} + q^2 C'_{33} - R \end{vmatrix} = 0. \qquad \text{(A6)}$$

This equation can be rewritten as

$$(C'_{66} - 2C'_{15}q + q^2 C'_{44} - R) \times$$

$$\left((C'_{11} + 2C'_{15}q + q^2 C'_{44} - R)(C'_{44} + q^2 C'_{33} - R) - (C'_{15} + q(C'_{13} + C'_{44}))^2\right) = 0, \qquad \text{(A7)}$$

In this equation, the first factor $(C'_{66} - 2C'_{15}q + q^2 C'_{44} - R)$ gives a solution for a shear wave propagation along the $x_2$ direction, while the second factor corresponds to the Rayleigh wave solution with motion in $(x_1, x_3)$ plane:

$$(C'_{11} + 2C'_{15}q + q^2 C'_{44} - R)(C'_{44} + q^2 C'_{33} - R) - \left(C'_{15} + q(C'_{13} + C'_{44})\right)^2 = 0. \qquad \text{(A8)}$$

This equation can be written as a fourth-order polynomial equation in terms of $q$,

$$aq^4 + bq^3 + cq^2 + dq + e = 0, \qquad \text{(A9)}$$

where $a = C'_{33}C'_{44}$, $b = 2C'_{15}C'_{33}$, $c = C'_{11}C'_{33} - {C'_{13}}^2 - 2C'_{13}C'_{44} - R(C'_{33} + C'_{44})$,
$d = -2C'_{15}(C'_{13} + R)$, and $e = C'_{11}C'_{44} - {C'_{15}}^2 - (C'_{11} + C'_{44})R + R^2$.

Equation (A9) has four generally complex roots. For a solid occupying the half-space $x_3 > 0$, the two roots with positive imaginary parts are selected to ensure decay of the wave amplitude with increasing depth. The displacement field can then be expressed as a superposition of the two corresponding solutions:

$$u_j = \sum_{n=1}^{2} A_j^{(n)} C_n \exp[ik \cdot (x_1 + q_n x_3 - ct)]\,. \qquad \text{(A10)}$$

Here, $A_2^1$ and $A_2^2$ are zero for a Rayleigh wave solution, as there is no displacement in $x_2$ direction. After defining the overall wave amplitude as $A_0$ and setting $C_1 = 1$, the parameter of the equation can be expressed as $A_1^1 = A_1^2 = A_0$, $A_3^1 = \gamma_1 A_0$, $A_3^2 = \gamma_2 A_0$, and $C_2 = C$. For the SAW propagation on the $(111)$ plane along the $[11\bar{2}]$ direction, $\gamma_n$ can be expressed as:

$$\gamma_n = -\frac{C'_{15}+(C'_{13}+C'_{44})q_n}{C'_{44}+C'_{33}(q_n)^2-R}. \tag{A11}$$

The two constants remaining unknown, $R$ and $C$, can be found from the free surface boundary condition,

$$\sigma_{i3} = C'_{i3lm} \frac{\partial u_m}{\partial x_l}\bigg|_{x_3=0} = 0. \tag{A12}$$

From $\sigma_{33}|_{x_3=0} = 0$, the parameter $C$ can be found as

$$C = -\frac{C'_{13}+C'_{33}\gamma_1 q_1}{C'_{13}+C'_{33}\gamma_2 q_2}, \tag{A13}$$

and from $\sigma_{13}|_{x_3=0} = 0$, an equation for finding the value of $R$ (and the speed $c$ of the SAW) can be obtained:

$$C'_{15}(1+C) + C'_{44}(\gamma_1 + \gamma_2 C + q_1 + q_2 C) = 0. \tag{A14}$$

The equations above determine all parameters required to evaluate Eq. (A10) and define the initial displacements, $u_j$, and velocities, $v_j = \partial u_j/\partial t$, for the MD simulation of a SAW propagating along the $[11\bar{2}]$ direction on the $(111)$ surface. The parameters calculated using the elastic constants of Pt predicted by the EAM potential at 300 K (Section 2) are listed in Table A1, together with the corresponding parameters for a SAW propagating along the $[100]$ direction on the $(001)$ surface calculated based on the solution provided in Ref. [33].

| Orientation | $\gamma_1$ | $\gamma_2$ | $C$ | $q_1$ | $q_2$ | $c\ [m/s]$ |
|---|---|---|---|---|---|---|
| $(111)[11\bar{2}]$ | $0.152681 + 0.327623i$ | $-5.1719 + 1.80706i$ | $-0.158447 + 0.395216i$ | $-0.699973 + 1.7386i$ | $0.115841 + 0.0789983i$ | 1278.03 |
| $(001)[100]$ | $-0.207492 + 1.13151i$ | $-0.207492 - 1.13151i$ | $-0.180368 + 0.983599i$ | $0.215003 + 0.779447i$ | $0.215003 - 0.779447i$ | 1316.02 |

**Table A1.** Parameters of Eq. (A10) for SAWs propagating along the $[11\bar{2}]$ direction on the $(111)$ surface and along the $[100]$ direction on the $(001)$ surface. The parameters are calculated using the elastic constants of Pt predicted by the EAM potential at 300 K and reported in Section 2.

## Supplementary Material

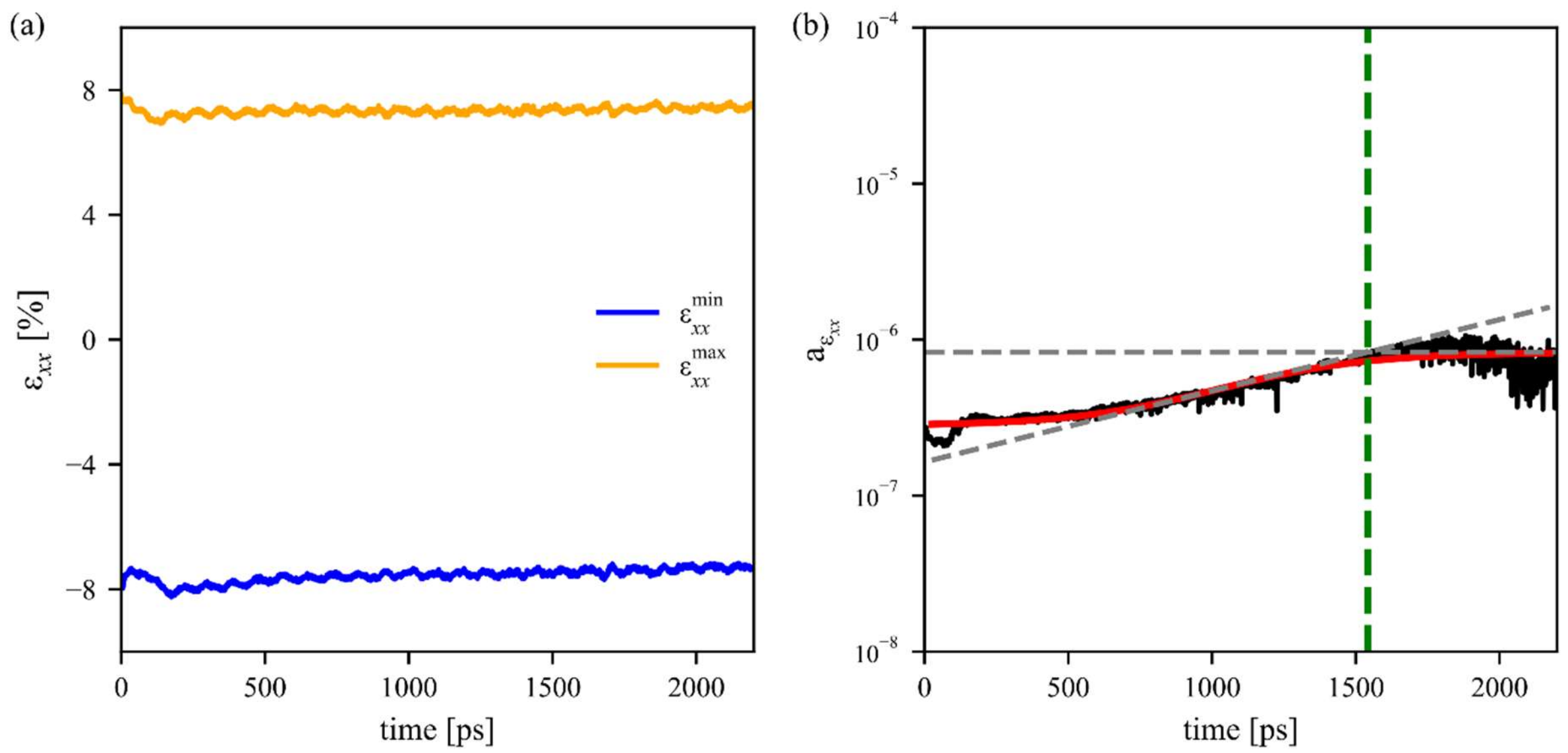

**Fig. S1.** Quantitative characterization of the evolution of surface strain profiles for a SAW with an initial strain amplitude of $\varepsilon_{xx}^{0} = 8\%$ propagating along the [100] direction on the (001) surface. (a) Evolution of the maximum and minimum values of surface strain $\varepsilon_{xx}$. (b) Local curvature $a_{\varepsilon_{xx}}$ obtained from a parabolic fit near the strain minimum, where a cusp develops during shock front formation. The red curve in (b) shows a sigmoid fit to the curvature data over the time interval from 25 to 2175 ps. The gray dashed lines indicate the tangent at the midpoint of the sigmoid and its saturated value; their intersection defines the shock formation time $\tau_{sh}$, marked by the green vertical dashed line.

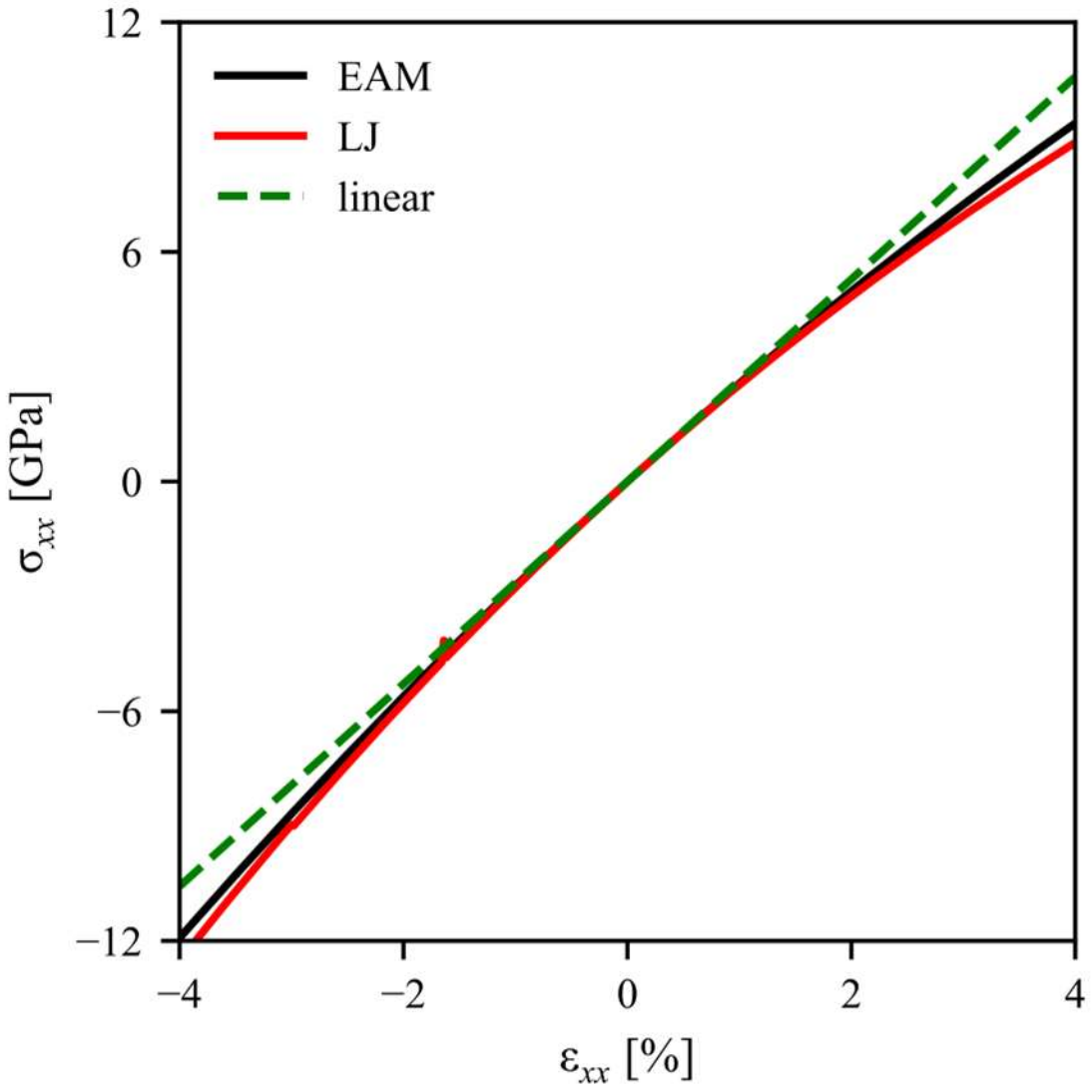

**Fig. S2.** Stress–strain response of Pt calculated using the EAM and LJ potentials at 0 K. Uniaxial tensile and compressive strains are applied along the *x* direction, with the dimensions along *y* and *z* held fixed. The simulations are performed for a cubic sample containing 25 conventional fcc unit cells along each of the *x* [100], *y* [010], and *z* [001] directions. Under these conditions, the slope of the stress–strain curve at zero strain corresponds to the elastic stiffness $C_{11}$. The green dashed line represents the linear elastic response calculated using the EAM value of $C_{11}$. The LJ parameters ε and σ are fitted to reproduce the equilibrium lattice constant and $C_{11}$ predicted by the EAM potential at 0 K.

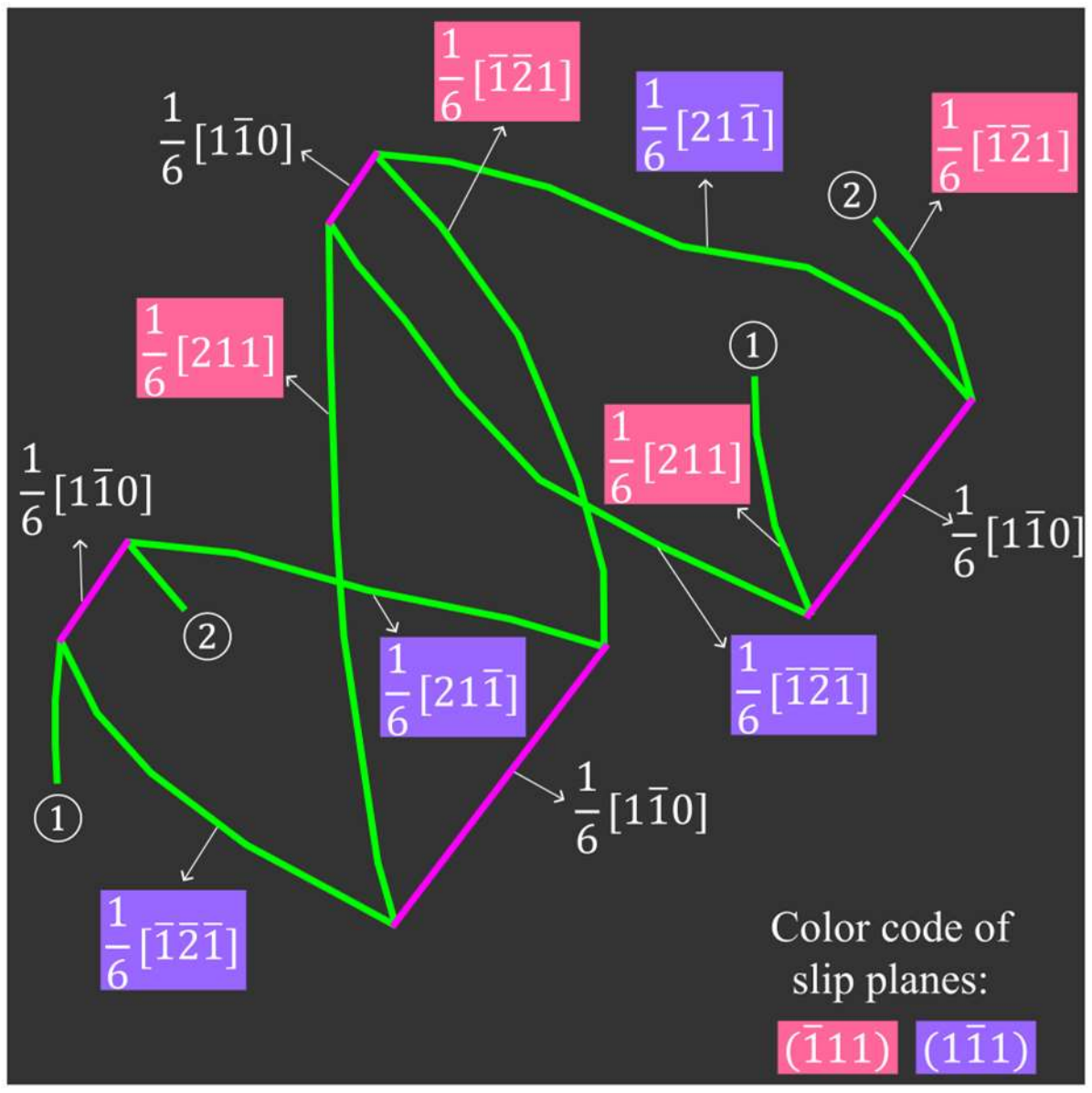

**Fig. S3.** Schematic representation of the dislocation structure outlined by box **C** in Fig. 16. This structure is a by-product of the rapid evolution of dislocation half-loops illustrated in Figs. 13-16. Dislocation lines are color-coded according to their types, as defined in the legend of Fig. 13. The circled numbers 1 and 2 indicate the points where the dislocation lines cross the periodic boundaries in the y direction. The Burgers vector of each dislocation is shown in a box with a background color identifying the corresponding slip plane, according to the color scheme defined in this figure.

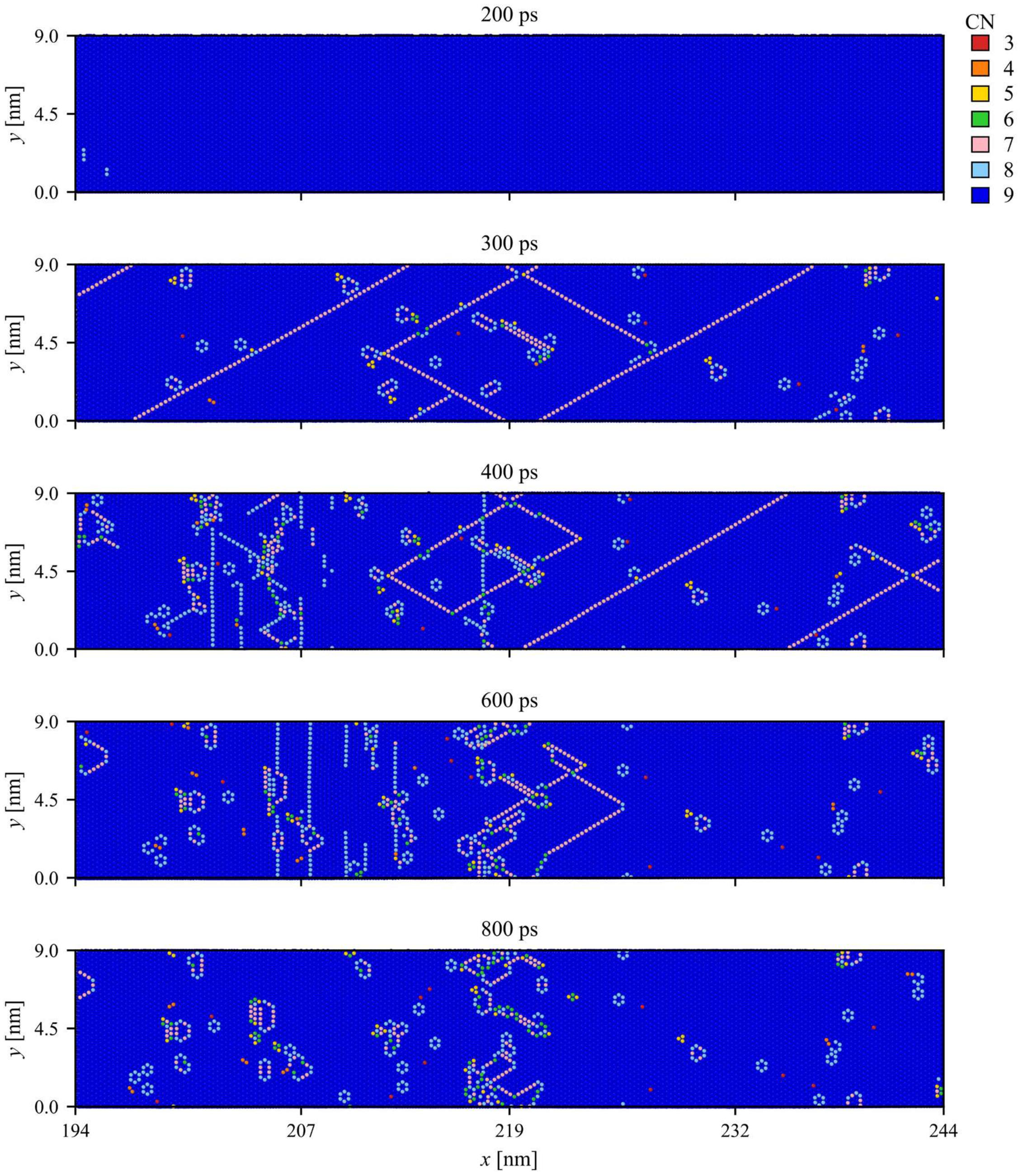

**Fig. S4.** Top views of a 50 nm long segment of the Pt(111) surface along the SAW propagation direction at selected times during the propagation of a SAW with an initial strain amplitude of $\varepsilon_{xx}^{0} = 4\%$. Atoms are colored according to their coordination numbers, using the same color scheme as in Fig. 22.

## References

[1] Lord Rayleigh, On waves propagated along a plane surface of an elastic solid, in: M. A. Pelissier, H. Hoeber, N. van de Coevering, I. F. Jones (Eds.), *Classics of Elastic Wave Theory*, SEG Geophysics Reprint Series No. 24, Society of Exploration Geophysicists, Tusla, Oklahoma, U.S.A., 2007, pp. 243-250.

[2] I. A. Viktorov, *Rayleigh and Lamb Waves: Physical Theory and Applications*, Plenum Press, New York, 1967.

[3] P. Hess, Surface acoustic waves in materials science, Phys. Today 55 (2002) 42-48.

[4] A. Lomonosov, A. P. Mayer, P. Hess, Laser-based surface acoustic waves in materials science, in: M. Levy, H. Bass, R. Stern (Eds.), *Modern Acoustical Techniques for the Measurement of Mechanical Properties*, Experimental Methods in the Physical Sciences, Vol. 39, Academic Press, San Diego, California, 2001, pp. 65-134.

[5] D. Schneider, Laser-induced surface acoustic waves for material testing, in: N. Ida, N. Meyendorf (Eds.), *Handbook of Advanced Non-Destructive Evaluation*, Springer International Publishing AG, Cham, Switzerland, 2018, pp. 1-64.

[6] K. H. Matlack, J.-Y. Kim, L. J. Jacobs, J. Qu, Review of second harmonic generation measurement techniques for material state determination in metals, J. Nondestruct. Eval. 34, (2015) 273.

[7] J. A. Cooper, R. A. Crosbie, R. J. Dewhurst, A. D. W. McKie, S. B. Palmer, Surface acoustic wave interactions with cracks and slots: A noncontacting study using lasers, IEEE Trans. Ultrason. Ferroelectr. Freq. Control. 33 (1986) 462-470.

[8] D. Chen, G. Lv, S. Guo, R. Zuo, Y. Liu, K. Zhang, Z. Su, W. Feng, Subsurface defect detection using phase evolution of line laser-generated Rayleigh waves, Opt. Laser Technol. 131 (2020) 106410.

[9] P. Dryburgh, W. Li, D. Pieris, R. Fuentes-Domínguez, R. Patel, R. J. Smith, M. Clark, Measurement of the single crystal elasticity matrix of polycrystalline materials, Acta Materialia 225 (2022) 117551.

[10] C. A. Dennett, K. P. So, A. Kushima, D. L. Buller, K. Hattar, M. P. Short, Detecting self-ion irradiation-induced void swelling in pure copper using transient grating spectroscopy, Acta Mater. 145 (2018) 496-503.

[11] F. Hofmann, M. P. Short, C. A. Dennett, Transient grating spectroscopy: An ultrarapid, nondestructive materials evaluation technique, MRS Bull. 44 (2019) 392-402.

[12] E. Botica-Artalejo, G. Wallace, M. P. Short, Real-time vacancy concentration evolution revealed via heavy ion irradiation experiments, arXiv:2602.18608, https://doi.org/10.48550/arXiv.2602.18608

[13] A. Reza, G. He, C. A. Dennett, H. Yu, K. Mizohata, F. Hofmann, Thermal diffusivity recovery and defect annealing kinetics of self-ion implanted tungsten probed by insitu transient grating spectroscopy, Acta Mater. 232 (2022) 117926.

[14] X. Ding, S.-C. S. Lin, B. Kiraly, H. Yue, S. Li, I-K. Chiang, J. Shi, S. J. Benkovic, T. J. Huang, On-chip manipulation of single microparticles, cells, and organisms using surface acoustic waves, Proc. Natl. Acad. Sci. U.S.A. 109 (2012) 11105-11109.

[15] M. C. Jo, R. Guldiken, Dual surface acoustic wave-based active mixing in a microfluidic channel, Sens. Actuators A Phys. 196 (2013) 1-7.

[16] L. Y. Yeo, J. R. Friend, Surface acoustic wave microfluidics, Annu. Rev. Fluid Mech. 46 (2014) 379-406.

[17] S. R. Heron, R. Wilson, S. A. Shaffer, D. R. Goodlett, J. M. Cooper, Surface acoustic wave nebulization of peptides as a microfluidic interface for mass spectrometry, Anal. Chem. 82 (2010) 3985-3989.

[18] J. Ho, M. K. Tan, D. B. Go, L. Y. Yeo, J. R. Friend, H.-C. Chang, Paper-based microfluidic surface acoustic wave sample delivery and ionization source for rapid and sensitive ambient mass spectrometry, Anal. Chem. 83 (2011) 3260-3266.

[19] Y. Huang, S. H. Yoon, S. R. Heron, C. D. Masselon, J. S. Edgar, F. Tureček, D. R. Goodlett, Surface acoustic wave nebulization produces ions with lower internal energy than electrospray ionization, J. Am. Soc. Mass Spectrom. 23 (2012) 1062-1070.

[20] A. A. Kolomenskii, H. A. Schuessler, V. G. Mikhalevich, A. A. Maznev, Interaction of laser-generated surface acoustic pulses with fine particles: Surface cleaning and adhesion studies, J. Appl. Phys. 84 (1998) 2404-2410.

[21] S. Arif, O. Armbruster, W. Kautek, Pulse laser particulate separation from polycarbonate: Surface acoustic wave and thermomechanical mechanisms, Appl. Phys. A 111 (2013) 539-548.

[22] C. Wu, V. Yu. Zaitsev, L. V. Zhigilei, Acoustic enhancement of surface diffusion, J. Phys. Chem. C 117 (2013) 9252-9258.

[23] M. V. Shugaev, A. J. Manzo, C. Wu, V. Yu. Zaitsev, H. Helvajian, L. V. Zhigilei, Strong enhancement of surface diffusion by nonlinear surface acoustic waves, Phys. Rev. B 91 (2015) 235450.

[24] A. J. Manzo, H. Helvajian, Demonstration of enhanced surface mobility of adsorbate cluster species by surface acoustic wave excitation induced by a pulsed laser, Proc. SPIE 8969 (2014) 896908.

[25] L. V. Zhigilei, H. Helvajian, Acoustic processes in materials, MRS Bull. 44 (2019) 345-349.

[26] S. P. Selvin, M. Esfandyarpour, A. Ji, Y. J. Lee, C. Yule, J.-H. Song, M. Taghinejad, M. L. Brongersma, Acoustic wave modulation of gap plasmon cavities, Science 389 (2025) 516-520.

[27] J. K. Wuenschell, H. Helvajian, Enhanced laser crystallization of thin film amorphous molybdenum disulfide ($MoS_2$) by means of pulsed laser ultrasound, Opt. Express 27 (2019) 5859-5873.

[28] D. Veysset, A. A. Maznev, I. A. Veres, T. Pezeril, S. E. Kooi, A. M. Lomonosov, K. A. Nelson, Acoustical breakdown of materials by focusing of laser-generated Rayleigh surface waves, Appl. Phys. Lett. 111 (2017) 031901.

[29] D. Veysset, S. E. Kooi, R. Haferssas, M. Hassani-Gangaraj, M. Islam, A. A. Maznev, Y. Chernukha, X. Zhao, K. Nakagawa, D. Martynowych, X. Zhang, A. M. Lomonosov, C. A. Schuh, R. Radovitzky, T. Pezeril, K. A. Nelson, Glass fracture by focusing of laser-generated nanosecond surface acoustic waves, Scr. Mater. 158 (2019) 42-45.

[30] A. M. Lomonosov, P. Hess, Impulsive fracture of silicon by elastic surface pulses with shocks, Phys. Rev. Lett. 89 (2002) 095501.

[31] V. V. Kozhushko, A. M. Lomonosov, P. Hess, Intrinsic strength of silicon crystals in pure- and combined-mode fracture without precrack, Phys. Rev. Lett. 98 (2007) 195505.

[32] G. Lehmann, A. M. Lomonosov, P. Hess, P. Gumbsch, Impulsive fracture of fused quartz and silicon crystals by nonlinear surface acoustic waves, J. Appl. Phys. 94 (2003) 2907-2914.

[33] M. V. Shugaev, C. Wu, V. Yu. Zaitsev, L. V. Zhigilei, Molecular dynamics modeling of nonlinear propagation of surface acoustic waves, J. Appl. Phys. 128 (2020) 045117.

[34] Y. Inoue, Y. Matsukawa, K. Sato, Effect of surface acoustic wave generated on ferroelectric support upon catalysis, J. Am. Chem. Soc. 111 (1989) 8965-8966.

[35] S. Kelling, T. Mitrelias, Y. Matsumoto, V. P. Ostanin, D. A. King, Acoustic wave enhancement of the catalytic oxidation of carbon monoxide over Pt {110}, J. Chem. Phys. 107 (1997) 5609-5612.

[36] S. Kelling, T. Mitrelias, Y. Matsumoto, V. P. Ostanin, D. A. King, Influence of acoustic wave excitation on CO oxidation over a Pt{110} single crystal, Faraday Discuss. 107 (1997) 435-444.

[37] S. Kelling, N. Saito, Y. Inoue, D. A. King, Surface morphological changes induced in catalysts by acoustic waves, Appl. Surf. Sci. 150 (1999) 47-57.

[38] Y. Inoue, Effects of acoustic waves-induced dynamic lattice distortion on catalytic and adsorptive properties of metal, alloy and metal oxide surfaces, Surf. Sci. Rep. 62 (2007) 305-336.

[39] Y. Inoue, Acoustic enhancement of surface reactions, MRS Bull. 44 (2019) 361-371.

[40] B. von Boehn, M. Foerster, M. von Boehn, J. Prat, F. Macià, B. Casals, M. W. Khaliq, A. Hernández-Mínguez, L. Aballe, R. Imbihl, On the promotion of catalytic reactions by surface acoustic waves, Angew. Chem. Int. Ed. 59 (2020) 20224-20229.

[41] E. A. Zabolotskaya, Nonlinear propagation of plane and circular Rayleigh waves in isotropic solids, J. Acoust. Soc. Am. 91 (1992) 2569-2575.

[42] A. Lomonosov, V. G. Mikhalevich, P. Hess, E. Y. Knight, M. F. Hamilton, E. A. Zabolotskaya, Laser-generated nonlinear Rayleigh waves with shocks, J. Acoust. Soc. Am. 105 (1999) 2093-2096.

[43] M. F. Hamilton, Yu. A. Il'inskii, E. A. Zabolotskaya, Nonlinear surface acoustic waves in crystals, J. Acoust. Soc. Am. 105 (1999) 639-651.

[44] Z. Liu, B. Lin, X. Liang, A. Du, X. Ma, Fracture of single crystal silicon caused by nonlinear evolution of surface acoustic waves, Eng. Fract. Mech. 269 (2022) 108505.

[45] Y. Wan, B. Lin, Z. Liu, X. Ma, H. Jia, W. Chen, Y. Li, Nonlinear surface acoustic wave characterization of single crystal copper: A molecular dynamics study, Mater. Today Commun. 40 (2024) 110053.

[46] C. J. O'Brien, C. M. Barr, P. M. Price, K. Hattar, S. M. Foiles, Grain boundary phase transformations in PtAu and relevance to thermal stabilization of bulk nanocrystalline metals, J. Mater. Sci. 53, (2018) 2911-2927.

[47] G. Clavier, N. Desbiens, E. Bourasseau, V. Lachet, N. Brusselle-Dupend, B. Rousseau, Computation of elastic constants of solids using molecular simulation: comparison of constant volume and constant pressure ensemble methods, Mol. Simul. 43 (2017) 1413-1422.

[48] A. P. Thompson, H. M. Aktulga, R. Berger, D. S. Bolintineanu, W. M. Brown, P. S. Crozier, P. J. in't Veld, A. Kohlmeyer, S. G. Moore, T. D. Nguyen, R. Shan, M. J. Stevens, J. Tranchida, C. Trott, S. J. Plimpton, LAMMPS - a flexible simulation tool for particle-based materials modeling at the atomic, meso, and continuum scales, Comput. Phys. Commun. 271 (2022) 108171.

[49] R. Stoneley, The propagation of surface elastic waves in a cubic crystal, Proc. R. Soc. London A 232 (1955) 447-458.

[50] A. P. Thompson, S. J. Plimpton, W. Mattson, General formulation of pressure and stress tensor for arbitrary many-body interaction potentials under periodic boundary conditions, J. Chem. Phys. 131 (2009) 154107.

[51] M. F. Hamilton, Y. A. Il'insky, E. A. Zabolotskaya, Local and nonlocal nonlinearity in Rayleigh waves, J. Acoust. Soc. Am. 97 (1995) 882-890.

[52] A. A. Kolomenskii, V. A. Lioubimov, S. N. Jerebtsov, H. A. Schuessler, Nonlinear surface acoustic wave pulses in solids: Laser excitation, propagation, interactions (invited), Rev. Sci. Instrum. 74 (2003) 448-452.

[53] Y. Xu, C. Chen, L. V. Zhigilei, Mechanisms and channels of energy dissipation in nonlinear surface acoustic wave propagation on a metal surface, in preparation.

[54] E. Y. Knight, M. F. Hamilton, Y. A. Il'inskii, E. A. Zabolotskaya, On Rayleigh wave nonlinearity, and analytical approximation of the shock formation distance, J. Acoust. Soc. Am. 102 (1997) 2529-2535.

[55] P. Virtanen, R. Gommers, T. E. Oliphant, M. Haberland, T. Reddy, D. Cournapeau, E. Burovski, P. Peterson, W. Weckesser, J. Bright, S. J. Van Der Walt, M. Brett, J. Wilson, K. J. Millman, N. Mayorov, A. R. J. Nelson, E. Jones, R. Kern, E. Larson, C. J. Carey, İ. Polat, Y. Feng, E. W. Moore, J. VanderPlas, D. Laxalde, J. Perktold, R. Cimrman, I. Henriksen, E. A. Quintero, C. R. Harris, A. M. Archibald, A. H. Ribeiro, F. Pedregosa, P. Van Mulbregt, SciPy 1.0 Contributors, SciPy 1.0: fundamental algorithms for scientific computing in Python, Nat. Methods 17 (2020) 261-272.

[56] L. K. Zarembo, V. A. Krasil'nikov, Nonlinear phenomena in the propagation of elastic waves in solids, Sov. Phys. Usp. 13 (1971) 778-797.

[57] Y. Zheng, R. Gr. Maev, I. Yu. Solodov, Nonlinear acoustic applications for material characterization: A review, Can. J. Phys. 77 (1999) 927-967.

[58] R. E. Kumon, M. F. Hamilton, Directional dependence of nonlinear surface acoustic waves in the (001) plane of cubic crystals, J. Acoust. Soc. Am. 111 (2002) 2060-2069.

[59] A. Stukowski, K. Albe, Extracting dislocations and non-dislocation crystal defects from atomistic simulation data, Modelling Simul. Mater. Sci. Eng. 18 (2010) 085001.

[60] A. Stukowski, Visualization and analysis of atomistic simulation data with OVITO–the Open Visualization Tool, Modelling Simul. Mater. Sci. Eng. 18 (2010) 015012.

[61] J. Frenkel, Zur Theorie der Elastizitätsgrenze und der Festigkeit kristallinischer Körper, Z. Phys. 37 (1926) 572-609.

[62] D. Roundy, C. R. Krenn, M. L. Cohen, J. W. Morris, Jr., Ideal shear strengths of fcc aluminum and copper, Phys. Rev. Lett. 82 (1999) 2713-2716.

[63] A. Kelly, N. H. MacMillan, *Strong Solids* (Oxford University Press, Oxford, 1986).

[64] A. T. Paxton, P. Gumbsch, M. Methfessel, A quantum mechanical calculation of the theoretical strength of metals, Phil. Mag. Lett. 63 (1991) 267-274.

[65] D. M. Clatterbuck, C. R. Krenn, M. L. Cohen, and J. W. Morris, Jr., Phonon instabilities and the ideal strength of aluminum, Phys. Rev. Lett. 91 (2003) 135501.

[66] S. Ogata, J. Li, S. Yip, Ideal pure shear strength of aluminum and copper, Science 298 (2002) 807-811.

[67] Y. Cai, H. A. Wu, S. N. Luo, A loading-dependent model of critical resolved shear stress, Int. J. Plast. 109 (2018) 1-17.

[68] T. Zhu, J. Li, Ultra-strength materials, Prog. Mater. Sci. 55 (2010) 710-757.

[69] S. Vlassov, B. Polyakov, L. M. Dorogin, M. Antsov, M. Mets, M. Umalas, R. Saar, R. Lõhmus, I. Kink, Elasticity and yield strength of pentagonal silver nanowires: In situ bending tests, Mater. Chem. Phys. 143 (2014) 1026-1031.

[70] L. Y. Chen, M. R. He, J. Shin, G. Richter, D. S. Gianola, Measuring surface dislocation nucleation in defect-scarce nanostructures, Nat. Mater. 14 (2015) 707-713.

[71] S.-H. Kim, J.-H. Seo, D.-M. Whang, J.-P. Ahn, J.-C. Lee, Deformation twinning of ultrahigh strength aluminum nanowire, Acta Mater. 160 (2018) 14-21.

[72] A. Gouldstone, H. J. Koh, K. Y. Zeng, A. E. Giannakopoulos, S. Suresh, Discrete and continuous deformation during nanoindentation of thin films, Acta Mater. 48 (2000) 2277-2295.

[73] T. Sato, Z. B. Milne, M. Nomura, N. Sasaki, R. W. Carpick, H. Fujita, Ultrahigh strength and shear-assisted separation of sliding nanocontacts studied in situ, Nat. Commun. 13 (2022) 2551.

[74] R. L. Fleischer, Cross slip of extended dislocations, Acta Metall. 7 (1959) 134-135.

[75] M. J. Buehler, A. Hartmaier, M. A. Duchaineau, F. F. Abraham, H. Gao, The dynamical complexity of work-hardening: a large-scale molecular dynamics simulation, Acta Mech. Sinica 21 (2005) 103-111.

[76] I. J. Beyerlein, A. Hunter, Understanding dislocation mechanics at the mesoscale using phase field dislocation dynamics, Phil. Trans. R. Soc. A 374 (2016) 20150166.

[77] M. Niewczas, Dislocations and twinning in face centred cubic crystals, Chapter 75 in: F. Nabarro, J. Hirth (Eds.), *Dislocations in Solids*, Elsevier Science, 2007, pp. 263-364.

[78] I. V. Ostrovskii, Ju. M. Khalack, A. B. Nadtochii, H. G. Walther, Defects modification by ultrasound in crystals, Solid State Phenom. 67-68 (1999) 497-502.

[79] A. A. Nazarov, M. A. Murzinova, A. A. Mukhametgalina, E. R. Shayakhmetova, Bulk ultrasonic treatment of crystalline materials, Metals 13 (2023) 344.

[80] Ch. Zaum, N. Osterloh, R. Darkins, D. M. Duffy, K. Morgenstern, Real-space observation of surface structuring induced by ultra-fast-laser illumination far below the melting threshold, Sci. Rep. 11 (2021) 13269.

[81] N. Osterloh, T. Pan, K. Morgenstern, Locally varying formation of nanoclusters across a low-intensity ultra-short laser spot, Nanoscale Horiz. 8 (2023) 55-62.

[82] N. Osterloh, M. Vyshnepolsky, T. Pan, and K. Morgenstern, Fluence-independent range of surface roughening within the Gaussian-shaped profile of a low-intensity ultrashort laser, Nanoscale Horiz. 11 (2026) 1676-1684.

[83] N. Tian, Z.-Y. Zhou, S.-G. Sun, Y. Ding, Z. L. Wang, Synthesis of tetrahexahedral platinum nanocrystals with high-index facets and high electro-oxidation activity, Science 316 (2007) 732-735.

[84] R. Narayanan, M. A. El-Sayed, Shape-dependent catalytic activity of platinum nanoparticles in colloidal solution, Nano Lett. 4 (2004) 1343-1348.

[85] F. Calle-Vallejo, J. Tymoczko, V. Čolić, Q. Huy Vu, M. D. Pohl, K. Morgenstern, D. Loffreda, P. Sautet, W. Schuhmann, A. S. Bandarenka, Finding optimal surface sites on heterogeneous catalysts by counting nearest neighbors, Science 350 (2015) 185-189.

[86] F. Calle-Vallejo, M. D. Pohl, D. Reinisch, D. Loffreda, P. Sautet, A. S. Bandarenka, Why conclusions from platinum model surfaces do not necessarily lead to enhanced nanoparticle catalysts for the oxygen reduction reaction, Chem. Sci. 8 (2017) 2283-2289.

[87] D. Royer, E. Dieulesaint, Rayleigh wave velocity and displacement in orthorhombic, tetragonal, hexagonal, and cubic crystals, J. Acoust. Soc. Am. 76 (1984) 1438-1444.